\documentclass[11pt]{article}

\usepackage{amssymb,multirow,color,comment,amsmath}
\usepackage{amsthm,tikz}
\usetikzlibrary{matrix}
\usepackage{pgfplots}
\usepackage{subfigure,bm}
\usepackage{hyperref}

\usetikzlibrary{arrows,calc}
\usepackage{multirow,comment,enumerate}
\usepackage{pgf,tikz,tikz-3dplot}
\usepackage{pdflscape}
\usepackage{pgfplots,xcolor} 
\usepackage{subfigure}
\usepackage{algorithm}
\usepackage{algpseudocode,mathtools}
\newcommand{\mc}{\multicolumn{1}{c}}
\newcommand{\tikzcircle}[2][black,fill=black]{\tikz[baseline=-0.5ex]\draw[#1,radius=#2] (0,0) circle ;}%
\newcommand{\tikzcirclehol}[2][black]{\tikz[baseline=-0.5ex]\draw[#1,radius=#2] (0,0) circle ;}%

\newcommand{\bE}{\mathbf{E}}
\newcommand{\br}{\mathbf{r}}
\newcommand{\sd}{\mbox{d}}
\newcommand{\te}{\mbox{e}}
\newcommand{\bP}{\mathbf{P}}
\newcommand{\bJ}{\mathbf{J}}

\newcommand{\matColumn}[1]{%
  \draw [fill = #1, draw, thin] (0, 0) rectangle (0.15, 0.15);
}%

\newcommand{\matDiag}[1]{%
  \draw [#1, draw, thin] (0, 0) rectangle (0.15, 0.15);
  \draw [#1] (0, 0.15) -- (0.15, 0);
}%

\newcommand{\matCol}[1]{%
  \draw [fill = #1, draw, thin] (0, 0) rectangle (0.09, 0.09);
}%

\newcommand{\matdiag}[1]{%
  \draw [#1, draw, thin] (0, 0) rectangle (0.09, 0.09);
  \draw [#1] (0, 0.09) -- (0.09, 0);
}%

\newcommand{\matcol}[1]{%
  \draw [fill = #1, draw, thin] (0, 0) rectangle (0.07, 0.07);
}%

\begin{document}

\title{Preconditioning the discrete dipole approximation}

\author{Samuel P. Groth$^{\text{a}}$\footnotemark[1]\footnote{Present address: Department of Engineering, University of Cambridge, United Kingdom.}\ \footnotemark[2]\footnote{Corresponding author. Email address: samuelpgroth@gmail.com}, Athanasios G. Polimeridis$^{\text{b}}$, and Jacob K. White$^{\text{a}}$\\[2pt]
$^{\text{a}}$Department of Electrical Engineering and Computer Science,\\ 
Massachusetts Institute of Technology, USA.\\[4pt]
$^{\text{b}}$Q Bio, Redwood, CA 94063, USA.
}

\maketitle

\begin{abstract}
The discrete dipole approximation (DDA) is a popular numerical method for calculating the scattering properties of atmospheric ice crystals. The standard DDA formulation involves the uniform discretization of the underlying volume integral equation, leading to a linear system with a block-Toeplitz Toeplitz-block matrix. This structure permits a matrix-vector product to be performed with $\mathcal{O}(n\log n)$ complexity via the fast-Fourier transform (FFT). Thus, in principle, the system can be solved rapidly using an iterative method. However, it is well known that the convergence of iterative methods becomes increasing slow as the optical size and refractive index of the scattering obstacle are increased. In this paper, we present a preconditioning strategy based on the multi-level circulant preconditioner of Chan and Olkin [Numer.~Algorithms \textbf{6}, 89 (1994)] and assess its performance for improving this rate of convergence. In particular, we approximate the system matrix by a block-circulant circulant-block matrix which can be inverted rapidly using the FFT. We present numerical results for scattering by hexagonal ice prisms demonstrating that this serves as an effective preconditioning strategy, reducing simulation times by orders of magnitude in many cases. A Matlab implementation of this work is freely available online.
\end{abstract}


\section{Introduction}
\label{sec:intro}
Since the publication of Purcell and Pennypacker's seminal paper in 1973~\cite{purcell1973scattering} and the subsequent work of, amongst others, Draine and Flatau~\cite{draine1994discrete}, Goedecke and O'Brien~\cite{goedecke1988scattering}, and Yurkin and Hoekstra~\cite{yurkin2007discrete}, the discrete dipole approximation (DDA) has proven popular for electromagnetic scattering simulations. Application areas include dust particles~\cite{purcell1973scattering,nousiainen2009single}, biological tissues~\cite{polimeridis2014stable,yurkin2005experimental}, optical tweezers~\cite{nieminen2007optical}, and atmospheric ice crystals~\cite{flatau2014light}.

The success of DDA is in no small part due to the fact that, when the underlying volume integral equation is discretized over a uniform (``voxelized'') grid, the system matrix obtains a block-Toeplitz Toeplitz-block (BTTB) structure. This permits a matrix-vector product to be performed in $\mathcal{O}(n\log n)$ operations via the fast-Fourier transform (FFT), where $n$ is the number of voxels in the grid~\cite{sarkar1986application,goodman1991application}. Therefore, the cost of solving the linear system via an iterative method is $\mathcal{O}(n_{\text{iter}} n\log n)$, where $n_{\text{iter}}$ is the number of iterations required for convergence. 

In principle, this modest growth of computational cost as $n$ increases should enable simulations for extremely large scattering obstacles to be efficiently performed. However, it is well known that, as the optical size and refractive index of the scatterer grow, $n_{\text{iter}}$ becomes prohibitively large (see, e.g., \cite{yurkin2007discrete}). In fact, as we shall observe later, $n_{\text{iter}}\sim x^3$, where $x=ka$ is the size parameter of the particle, with $k$ the wavenumber and $a$ the radius of the obstacle's smallest circumscribing sphere. In ice crystal simulations, for example, the scatterer may be up to a hundred wavelengths across for which $n_{\text{iter}}$ is so large as to make DDA infeasible. Therefore, an effective preconditioning strategy is required to temper the growth of $n_{\text{iter}}$. 

This paper revisits the well-established circulant preconditioning techniques of Chan and Olkin~\cite{chan1994circulant,chan1988optimal}, in which the underlying Toeplitz matrix is approximated by a circulant, and applies it within the DDA context for the first time. This builds on work by the present authors where a similar approach was applied within a silicon photonics context in which the structures of interest typically have extreme length in one of the three physical dimensions~\cite{groth2018circulant,tambova2018adiabatic,groth2019circulant}.

To clarify briefly this approach, consider the structure of the DDA linear system. It is of the form
\begin{equation}
	(\textbf{D} + \textbf{T})\textbf{x} = \textbf{b},
	\label{eqn:general_system}
\end{equation}
where $\textbf{D}$ is a diagonal matrix containing the polarizabilities of the dipoles and $\textbf{T}$ is a BTTB matrix with three levels of Toeplitz structure, obtained from discretizing the dyadic Green's function over the three-dimensional voxelized grid. For optically large scatterers, solving this system iteratively is expensive due to a large $n_{\text{iter}}$, hence we seek an appropriate preconditioner $\textbf{P}$ such that the modified system
\begin{equation}
	\textbf{P}^{-1}(\textbf{D} + \textbf{T})\textbf{x} = \textbf{P}^{-1}\textbf{b}
	\label{eqn:general_prec}
\end{equation}
is more efficiently solved, i.e., $n_{\text{iter}}$ is drastically reduced. 

In order for $\textbf{P}$ to be effective, it must also be reasonably cheap to construct and invert. In our Toeplitz setting, a natural candidate for $\textbf{P}$ is a circulant matrix. Circulant matrices constitute a special class of Toeplitz matrix with the additional desirable property that they are diagonalized by the discrete Fourier transform, and hence can be inverted in $\mathcal{O}(n\log n)$. Circulant preconditioners for Toeplitz systems have proven to be successful in many application areas (see \cite{ng2004iterative} and the references therein), however, to the present authors' knowledge, they have yet to be applied to DDA for practical EM scattering problems.

The shrewd reader would have observed that the matrix $\textbf{D}$ does not, in general, have a constant diagonal (only when the scatter is a homogeneous cuboid is this constant). Hence, $(\textbf{D}+\textbf{T})$ does not inherit the Toeplitz property of $\textbf{T}$ for general scatterers. To circumvent this issue, we employ a simple averaging strategy to create a preconditioner of the form $\textbf{P}=\tilde{\textbf{D}} + \textbf{C}$, where $\tilde{\textbf{D}}$ has constant diagonal and is constructed by averaging the diagonal of $\textbf{D}$, and $\textbf{C}$ is a circulant approximation to $\textbf{T}$ \`{a} la Chan and Olkin~\cite{chan1994circulant}. Since $\tilde{\textbf{D}}$ has constant diagonal, then $\tilde{\textbf{D}} + \textbf{C}$ inherits the circulant nature of $\textbf{C}$, and hence the preconditioner $\textbf{P}$ can be cheaply inverted. 

The example scatterers considered in this article are homogeneous hexagonal ice prisms of various size parameters and refractive indices. The discretized domain is the smallest box enclosing the scatterer (this is required for the FFT acceleration) so that the values along the diagonal of $\textbf{D}$ correspond to those of ice and air voxels. Here we have chosen to construct $\tilde{\textbf{D}}$ as $\tilde{\textbf{D}}=\text{mode}(\text{diag}(\textbf{D}))\textbf{I}$, where $\textbf{I}$ is the identity matrix, i.e., the constant diagonal of $\tilde{\textbf{D}}$ is the modal average of the diagonal entries in $\textbf{D}$. For the hexagonal prisms considered here, this means that the preconditioner corresponds to the total bounding box ``filled in'' with ice. As we see in Section~\ref{sec:num_results}, this is an effective strategy for the ice crystal examples considered in this article. Improvements may potentially be made by considering either a different averaging technique or a more sophisticated  ``Toeplitz-plus-diagonal'' preconditioner, as in \cite{ng2010approximate}, however these ideas are not explored here.

In terms of previous work on preconditioning for DDA, there appears to have been little development. Some brief experiments were presented in \cite{flatau1997improvements} where the simple point-Jacobi and Neumann polynomial preconditioners were used. However, large size parameters were not investigated and the improvement for small size parameters is extremely modest, if anything at all. More general preconditioning strategies exist, such as incomplete-LU, block-Jacobi~\cite{carpentieri2012fast}, and the inverse fast multipole method~\cite{coulier2017inverse}, but these are potentially expensive, are not effective for high frequency problems, or are complicated to implement. The distinct advantages of circulant preconditioning are that it is well suited to the Toeplitz structure of DDA, is comparatively straightforward to implement, and it is inexpensive. Furthermore, as we present in Section~\ref{sec:num_results}, circulant preconditioning performs extremely well for the ice crystal scattering simulations considered here, providing speed-up factors of ten or more, and in many cases it enables previously inaccessible simulations to be performed on a desktop PC. Therefore, we believe that this paper presents the first viable preconditioning approach for an important class of DDA scattering simulations. 

The layout of the paper is as follows. In Section~\ref{sec:IE}, we provide details of the standard DDA formulation of \cite{draine1994discrete} employed here. Also details of the Toeplitz structure of the system matrix are provided to facilitate the description of the circulant approximation in the following section. In Section~\ref{sec:prec}, we review circulant preconditioning applied to Toeplitz matrices and its extension to Toeplitz-block matrices. In particular, we describe how the general approach of \cite{chan1994circulant} is applied to our particular BTTB DDA matrix. Details of the algorithmic costings of assembling and applying the preconditioner are provided. We also present some pseudocode to help readers incorporate this preconditioning strategy into their own DDA codes. In Section~\ref{sec:num_results}, we consider the scattering of a polarized plane wave by hexagonal prisms of refractive indices $\mu =1.2,1.4,1.6,1.8,2$ and of size parameters $x = 10, 20, 30, 40, 60, 80, 100$. We present the CPU times and iteration counts for the unpreconditioned and preconditioned iterative solves of the arising linear systems, using both GMRES and BiCG-Stab. For smaller size parameters, little gain is achieved, but for large size parameters, we observe acceleration factors of ten or more. In fact, for the largest size parameters, where unpreconditioned DDA fails to converge, we achieve convergence with the preconditioned DDA within an acceptable number of iterations. In Section~\ref{sec:conc}, we provide some concluding remarks and ideas for the further development of the preconditioning strategy. 

A Matlab implementation is openly available online (https://github.com/samuelpgroth/VoxScatter) which we hope will be useful to students and those wishing to develop this work further.

\section{Integral equation formulation}
\label{sec:IE}
The discrete dipole approximation, in its many forms, begins with the following integral equation representation for the time-harmonic ($\te^{-i\omega t}$) electric field $\bE$ in the presence of a non-magnetic dielectric body $\Omega$:
\begin{equation}
	\bE(\br) = \bE^{\text{inc}}(\br) + \int_{\Omega} \mathbf{G}(\br,\br')\chi(\br')\bE(\br')\sd \br',
	\label{eqn:IE_rep}
\end{equation}
where $\bE^{\text{inc}}$ is the incident field and $\chi(\br) := (\epsilon(\br)-1)/4\pi$ is the electric susceptibility, with $\epsilon(\br)$ the relative permittivity. The dyadic Green's function, $\mathbf{G}$, is defined as
\begin{equation}
	\mathbf{G}(\br,\br') = (k_0^2\textbf{I} + \nabla\nabla)\frac{e^{ik_0r}}{r} = \frac{e^{ik_0r}}{r}\left[k_0^2\left(\mathbf{I}_3-\hat{\br}\hat{\br}^{\text{T}}\right)+\frac{ik_0r-1}{r^2}\left(\textbf{I}_3-3\hat{\br}\hat{\br}^{\text{T}}\right)\right],
\end{equation}
where $r=|\br-\br'|$, $\hat{\br} = (\br'-\br)/r \in \mathbb{R}^{3\ \times \ 1}$, and $\mathbf{I}$ is the $3\times~3$ identity matrix~\cite{draine1994discrete,yurkin2007discrete}.
Reordering (\ref{eqn:IE_rep}) to obtain an integral equation for the unknown field gives
\begin{equation}
	\left(\mathcal{I} - \mathcal{G}\chi\right)\bE = \bE^{\text{inc}},
	\label{eqn:IE}
\end{equation}
where $\mathcal{G}$ is the integral operator
\begin{equation}
	\mathcal{G}\mathbf{f}(\br) = \int_{\Omega} \mathbf{G}(\br,\br')\mathbf{f}(\br')\sd \br'.
\end{equation}

In the DDA of Purcell and Pennypacker~\cite{purcell1973scattering}, and Draine and Flatau~\cite{draine1994discrete}, equation (\ref{eqn:IE}) is phrased in terms of the unknown polarization rather than the electric field. The polarization is defined as
\begin{equation}
	\bP(\br) = \chi(\br)\bE(\br)
	\label{eqn:polarization}
\end{equation}
and so the integral equation becomes
\begin{equation}
	(\boldsymbol{\chi}^{-1} - \mathcal{G})\bP = \bE^{\text{inc}},
	\label{eqn:PVIE}
\end{equation}
for $\chi(\br)\neq0$ (of course $\mathbf{P}(\br)=0$ where $\chi(\br)=0$ so we can neglect the contributions from such voxels).
The formulation (\ref{eqn:PVIE}) is seen as desirable since there exists an accurate approximation for the self term via the Clausius-Mossotti relation. This enables the complicated evaluation of the singular portion of the integral to be sidestepped. 

In this paper, we solve (\ref{eqn:PVIE}) via the ``classical'' DDA approach as expounded in, for example, \cite{purcell1973scattering,draine1994discrete}. A more rigorous approach would be to solve (\ref{eqn:PVIE}) via Galerkin's method and evaluate the resulting double integrals with sophisticated numerical quadrature, as is done in \cite{polimeridis2014stable} where it is used for magnetic resonance applications. Here we choose to present results for the simpler DDA approach, but point out that the preconditioning strategy presented can be applied to any volume integral equation scheme (e.g., DDA, Galerkin, collocation). The only requirement is that a cuboidal discretization grid is used, so that the resulting linear system has Toeplitz structure.

\subsection{Discrete Dipole Approximation}

DDA can be viewed as a collocation approach for solving equation (\ref{eqn:IE}) in which the singular self-term integrals are evaluated using semi-analytical means (namely, the Claussius-Mossotti relation) and the non-singular integrals are evaluated using the midpoint quadrature rule. We briefly summarize this approach.

Begin by writing the unknown polarization $\mathbf{P}$ as 
\begin{equation}
	\mathbf{P}(\br) = \sum_{j=1}^N\mathbf{c}_j\circ\mathbf{p}_j(\br),
	\label{eqn:PWC}
\end{equation}
where each basis function $\mathbf{p}_j$ is a three-dimensional unit pulse function supported on voxel $j$ alone, i.e.,
\[
\mathbf{p}_j(\br) = 
	\begin{cases}
		(1,1,1),&\ \br \ \text{in voxel}\ j, \\
		0,  &\ \text{otherwise},
	\end{cases}
\]
$\mathbf{c}_j=(c_j^x,c_j^y,c_j^z)$ are the unknown coefficient vectors, and $\circ$ represents the Hadamard product. Upon substitution of the piecewise constant representation (\ref{eqn:PWC}) into the integral equation (\ref{eqn:PVIE}), and then forcing this to be exact at the voxel centers, we obtain the linear system of equations
\begin{equation}
	\sum_{j=1}^N \mathbf{c}_j\circ\left\{\chi^{-1}(\br_j)\mathbf{p}_j-\int_{\Omega_j}\mathbf{G}(\br_i,\br')\mathbf{p}_j(\br')\sd\br'\right\}= \bE^{\text{inc}}(\br_i),
	\label{eqn:lin_sys}
\end{equation}
for $i=1,\ldots,N$. This is the collocation approach for the solution of (\ref{eqn:PVIE}). We observe that when $i=j$ (the self term), the integral in (\ref{eqn:lin_sys}) is singular. In DDA schemes, this self term is given explicitly via the Clausius-Mossotti relation:
\begin{equation}
	\chi^{-1}(\br_i)\mathbf{p}_i-\int_{\Omega_i}\mathbf{G}(\br_i,\br')\mathbf{p}_i(\br')\sd\br' \approx \alpha_i^{-1}\mathbf{p}_i,\quad i=j,
	\label{eqn:CM}
\end{equation}
with $\alpha_i$ the polarizability of a dipole at location $\br_i$. Here we follow \cite{draine1994discrete} and take $\alpha = \alpha^{\text{LDR}}$, namely the lattice dispersion relation (LDR) correction to the Clausius-Mossotti polarizabilities $\alpha^{\text{CM}}$. The definitions are given as 
\begin{align*}
	&\alpha_i^{\text{LDR}} = \frac{\alpha_i^{\text{CM}}}{1+\alpha_i^{\text{CM}}[(b_1+m_i^2b_2+m_i^2b_3S)(k_0\Delta)^2-(2i/3)(k_0\Delta)^3]}, \\
	&\alpha_i^{\text{CM}} = \frac{3}{4\pi}\frac{\epsilon_i-1}{\epsilon_i+2}, \\
	& b_1 = -1.891531, \quad b_2=0.1648469, \\
	& b_3 = -1.7700004, \quad S:=\sum_{j=1}^3(d_j E_j)^2,
\end{align*}
where $\epsilon_i=\mu_i^2$ is the relative permittivity of the material occupying the $i$th voxel, and $\mathbf{d}^i=(d_1,d_2,d_3)$ and $\mathbf{E}_0=(E_1,E_2,E_3)$ are unit vectors defining the direction and polarization of the incident field. Note that we use slightly different definitions of $\alpha^{\text{CM}}$ and $\alpha^{\text{LDR}}$ to that used in \cite{draine1994discrete}, namely we omit the scaling by the voxel volume. But this difference is compensated for in our modified definition of the polarization (\ref{eqn:polarization}), which we choose to fall in line with the more standard definition.

The non-singular integrals ($i\neq j$) are evaluated using the midpoint quadrature rule:
\begin{equation}
	\int_{\Omega_j}\mathbf{G}(\br_i,\br')\mathbf{p}_j(\br')\sd\br' \approx \Delta^3 \mathbf{G}(\br_i,\br_j), \quad i\neq j,
	\label{eqn:midpoint}
\end{equation}
where $\Delta$ is the voxel dimension (see Section~\ref{ss:discretization}). Such a crude quadrature scheme is accurate only for well-separated voxels. For nearby voxels, where the integral is close to singular, the midpoint rule is inaccurate. Schemes such as the digitized Green's function~\cite{goedecke1988scattering}, coupled dipole method~\cite{chaumet2004coupled}, and the Galerkin implementation~\cite{polimeridis2014stable} use rigorous quadrature techniques to evaluate these integrals more accurately, and so are more accurate in general, and particularly for large permittivities. However, here we choose the more classical approach for simplicity. The important point is that both approaches are based on voxel discretizations, so can employ the preconditioning strategy proposed in this article.

\subsection{Voxel discretization and Toeplitz structure}
\label{ss:discretization}
Although tetrahedral discretizations (e.g., \cite{markkanen2016numerical}) can provide a more accurate geometrical representation, voxel discretizations have proven popular owing to the fact that they lead to a discrete system of convolution form, and hence permit a fast matrix-vector product via the FFT. 

We begin the discretization by choosing an appropriate voxel dimension $\Delta$. Typically $\Delta$ is chosen so that $\lambda/(\mu\Delta)\geq 10$ in order to ensure an accurate approximation, where $\lambda$ is the wavelength of the incident field. In this article, we take $\lambda/(\mu\Delta)= 10$ to enable rapid simulations with meaningful results.
Then a box bounding the scatterer is constructed, of dimension $l\Delta \times m\Delta \times n\Delta$ so that the voxel grid consists of $N = l\times m\times n$ voxels.

\begin{algorithm}
\caption{Computing the tensor $\textbf{G}\in\mathbb{C}^{l\ \times\ m\ \times\ n \ \times\ 6}$}
\label{array-sum}
\begin{algorithmic}[1]
\State $\br_0 = (x(1), y(1), z(1))\quad$ centerpoint of first voxel
\For {$i=1:l$}
    	\For {$j=1:m$}
		\For {$k=1:n$}
      			  \State $\br_c = (x(i), y(j), z(k))\quad$ centerpoint of voxel
			  \State $r = |\br_c - \br_0|$ 
			  \If {$r\neq 0$}
				  \State $\hat{\br} = (\br_c-\br_0)/r$ 
				  \State $\mathbf{g} = \frac{e^{ik_0r}}{r}\left[k_0^2(\mathbf{I}_3 - \hat{\br}^{\text{T}}\hat{\br}) + \frac{ik_0r-1}{r^2}(\mathbf{I}_3 - 3\hat{\br}^{\text{T}}\hat{\br})\right]$
				  \State $\mathbf{G}(i,j,k,1) = \mathbf{g}(1,1), \mathbf{G}(i,j,k,2) = \mathbf{g}(1,2)$
				  \State $\mathbf{G}(i,j,k,3) = \mathbf{g}(1,3), \mathbf{G}(i,j,k,4) = \mathbf{g}(2,2)$
				  \State $\mathbf{G}(i,j,k,5) = \mathbf{g}(2,3), \mathbf{G}(i,j,k,6) = \mathbf{g}(3,3)$
			\EndIf
		\EndFor
	\EndFor
\EndFor
\end{algorithmic}
\label{alg:G}
\end{algorithm}

Discretizing the linear system of equations (\ref{eqn:lin_sys}) over the voxel grid using (\ref{eqn:CM}) and (\ref{eqn:midpoint}), and using the ordering described in Algorithm~\ref{alg:G} leads to a discrete system of the form
\begin{equation*}
\hspace*{-0.5cm}
\left[
 \left(
  \begin{array}{ c c c | c c c | c c c }
     & &  &  &  & \mc{}& & &  \\
     & &  &  &  & \mc{} & & &  \\
    \multicolumn{3}{c|}{\raisebox{1\normalbaselineskip}[0pt][0pt]{$\boldsymbol{\alpha}_{x}^{-1}$}} &  \multicolumn{3}{c}{} &  \multicolumn{3}{c}{}  \\
    \cline{1-6}
      & &  &  &  & & & &   \\
      & &  &  &  & & & &  \\
     \multicolumn{3}{c|}{} & \multicolumn{3}{c|}{\raisebox{1\normalbaselineskip}[0pt][0pt]{$\boldsymbol{\alpha}_{y}^{-1}$}} &  \multicolumn{3}{c}{} \\
    \cline{4-9}
     & & \mc{} &  &  & & & &  \\
     & &\mc{}  &  &  & & & &  \\
     \multicolumn{3}{c}{} &  \multicolumn{3}{c|}{}& \multicolumn{3}{c}{\raisebox{1\normalbaselineskip}[0pt][0pt]{$\boldsymbol{\alpha}_{z}^{-1}$}}
  \end{array}
  \right) 
  -
  \renewcommand{\arraystretch}{0.9}
  \Delta^3\left(
  \begin{array}{ c c c | c c c | c c c }
     & &  &  &  & & & &  \\
     & &  &  &  & & & &  \\
    \multicolumn{3}{c|}{\raisebox{1\normalbaselineskip}[0pt][0pt]{$\textbf{G}_{xx}$}} &  \multicolumn{3}{c|}{\raisebox{1\normalbaselineskip}[0pt][0pt]{$\textbf{G}_{xy}$}} &  \multicolumn{3}{c}{\raisebox{1\normalbaselineskip}[0pt][0pt]{$\textbf{G}_{xz}$}}  \\
    \cline{1-9}
      & &  &  &  & & & &   \\
      & &  &  &  & & & &  \\
     \multicolumn{3}{c|}{\raisebox{1\normalbaselineskip}[0pt][0pt]{$\textbf{G}_{xy}$}} & \multicolumn{3}{c|}{\raisebox{1\normalbaselineskip}[0pt][0pt]{$\textbf{G}_{yy}$}} &  \multicolumn{3}{c}{\raisebox{1\normalbaselineskip}[0pt][0pt]{$\textbf{G}_{yz}$}} \\
    \cline{1-9}
     & &  &  &  & & & &  \\
     & &  &  &  & & & &  \\
     \multicolumn{3}{c|}{\raisebox{1\normalbaselineskip}[0pt][0pt]{$\textbf{G}_{xz}$}} &  \multicolumn{3}{c|}{\raisebox{1\normalbaselineskip}[0pt][0pt]{$\textbf{G}_{yz}$}}& \multicolumn{3}{c}{\raisebox{1\normalbaselineskip}[0pt][0pt]{$\textbf{G}_{zz}$}}
  \end{array}
  \right) 
  \right]
    \renewcommand{\arraystretch}{0.9}
    \left(
  \begin{array}{c}
   \\
   \\
  \raisebox{1\normalbaselineskip}[0pt][0pt]{$\textbf{c}_{x}$} \\
  \hline
  \\
   \\
  \raisebox{1\normalbaselineskip}[0pt][0pt]{$\textbf{c}_{y}$} \\
  \hline
  \\
   \\
  \raisebox{1\normalbaselineskip}[0pt][0pt]{$\textbf{c}_{z}$} 
  \end{array}
\right) 
=
  \renewcommand{\arraystretch}{0.9}
 \left(
  \begin{array}{c}
   \\
   \\
  \raisebox{1\normalbaselineskip}[0pt][0pt]{$\textbf{E}^{\text{inc}}_x$} \\
  \hline
  \\
   \\
  \raisebox{1\normalbaselineskip}[0pt][0pt]{$\textbf{E}^{\text{inc}}_y$} \\
  \hline
  \\
   \\
  \raisebox{1\normalbaselineskip}[0pt][0pt]{$\textbf{E}^{\text{inc}}_z$} 
  \end{array}
\right).
\label{eqn:sys_mat}
\end{equation*}
\normalsize
The blocks $\boldsymbol{\alpha}_{x}^{-1},\ \boldsymbol{\alpha}_y^{-1},\ \boldsymbol{\alpha}_z^{-1}$ are diagonal and each of the blocks $\textbf{G}_{\alpha\beta}$ has BTTB structure on three levels, corresponding to the three physical dimensions of the problem. Note the symmetry in these blocks: only six of them are unique. Further, each of these blocks is either symmetric or anti-symmetric. This, combined with their BTTB structure, allows them to be each defined by a single row. Hence the storage cost for the $\textbf{G}$ matrix is $\mathcal{O}(6n)$.

Further note that if the matrix $\boldsymbol{\alpha}$ has a constant diagonal, i.e., the structure is homogeneous, then the matrix $\boldsymbol{\alpha}^{-1}-\textbf{G}$ inherits the BTTB structure of $\textbf{G}$. This is the particular case in which circulant preconditioners prove most effective, as we discuss in the following section.

\section{Circulant preconditioning}
\label{sec:prec}
The circulant preconditioners employed here are based on those proposed in \cite{chan1994circulant} for Toeplitz-block matrices, which are an extension of the optimal point-circulant preconditioners of \cite{chan1988optimal}. We review here the salient features of multi-level circulant preconditioners and refer the reader to \cite{chan1994circulant} for further details.

A Toeplitz matrix $\text{T}_n = \{t_{ij}\}_{i,j=0}^{l-1}$ is Toeplitz if $t_{ij}=t_{i-j}$, i.e., the diagonals are constant. Circulant matrices $\text{C}_l=[c_{ij}]_{i,j=0}^{l-1}$ are also Toeplitz but with the additional property that every row of the matrix is a right cyclic shift of the row above, i.e, $c_{ij}=c_{(i-j)\ \text{mod}\ l}$.  Written out, these matrices have the respective forms
\small
\begin{equation*}
\text{T}_n = 
\left( \begin{array}{cccc}
t_0    & t_{-1}  & \ldots &   t_{-(l-1)} \\
t_1    & t_0    &     \ddots       &     \vdots \\
\vdots&   \ddots       &   \ddots    & t_{-1}  \\
t_{l-1}  &  \ldots    & t_1 & t_0   
\end{array} \right), \quad 
\text{C}_n = 
\left( \begin{array}{cccc}
c_0    & c_{l-1}   & \ldots & c_{1} \\
c_1    & c_0 &   \ddots & \vdots\\
\vdots &   \ddots     &   \ddots  &  c_{l-1}   \\
c_{l-1}  &  \ldots & c_1     & c_0
\end{array} \right).
\end{equation*}
\normalsize
Note that circulant matrices have the desirable property that they are diagonalized by the discrete Fourier matrix $\text{F}_l$, such that $\text{C}_l = \text{F}_l^{-1} \Lambda_l\text{F}_l$, where $\Lambda_l=\text{diag}(\text{F}_l\mathbf{c})$ is a diagonal matrix with $\mathbf{c}$ the defining column of $\text{C}_l$. Therefore, $\text{C}_l$ is inverted via the FFT in $\mathcal{O}(l\log l)$ operations. For a Toeplitz matrix, T.~Chan \cite{chan1988optimal} proposed the \textit{optimal} point-circulant preconditioner whose entries are given by
\begin{equation}
			c_i = 
            \begin{cases}
            				\frac{l-i}{l}t_i + \frac{i}{l}t_{-(l-i)}, & 0\leq i \leq l-1, \\
                            c_{l+i}, 		& -(l-1)\leq i <0.
              \end{cases}
              \label{eqn:Chan}
\end{equation}
This approximation is optimal in the sense that it is the closest circulant matrix to $\text{T}_l$ in the Frobenius norm. There exist other circulant preconditioners (see, for example, the review \cite{strela1996circulant}) and we anticipate the results presented in this paper would be similar if, for example, the Strang circulant preconditioner~\cite{strang1986proposal} were instead employed. We choose to employ T.\ Chan's preconditioner since it is explicitly defined by the simple formula (\ref{eqn:Chan}) and has been shown to be effective for many Toeplitz problems.

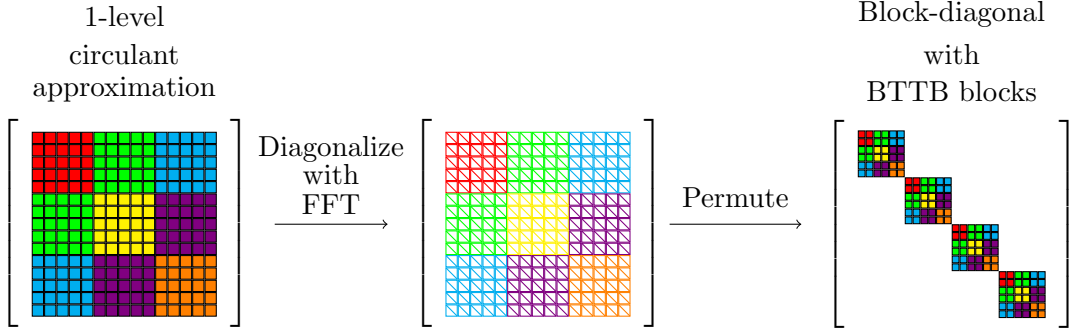
\begin{figure*}[t!]
\centering

\begin{tikzpicture}[
  node distance = 0.3cm,
  mtn/.style = {
    matrix of nodes,
    left delimiter = {[},
    right delimiter = {]}
  },
  ]

  \draw(0,2.5) node[anchor=south]{1-level};
    \draw(0, 2) node[anchor=south]{circulant};
    \draw(0,1.5) node[anchor=south]{approximation};
  \begin{scope}
   \matrix[mtn] (Y) {%
   \matColumn{red}; &  \matColumn{red}; &  \matColumn{red};& \matColumn{red}; &\matColumn{red};  &\matColumn{green};  &\matColumn{green};  &\matColumn{green};  &\matColumn{green};  & \matColumn{green};   &\matColumn{cyan};   & \matColumn{cyan};  &  \matColumn{cyan}; & \matColumn{cyan};  &  \matColumn{cyan};    \\ 
 \matColumn{red};   & \matColumn{red}; &  \matColumn{red};  &\matColumn{red};   &\matColumn{red};   &\matColumn{green};   &\matColumn{green};   &\matColumn{green};   &\matColumn{green};   & \matColumn{green};  & \matColumn{cyan};  & \matColumn{cyan};  & \matColumn{cyan};  & \matColumn{cyan};  & \matColumn{cyan};       \\ 
 \matColumn{red};   & \matColumn{red}; &  \matColumn{red};  &\matColumn{red};   &\matColumn{red};   &\matColumn{green};   &\matColumn{green};   &\matColumn{green};   &\matColumn{green};   & \matColumn{green};  & \matColumn{cyan};  & \matColumn{cyan};  & \matColumn{cyan};  & \matColumn{cyan};  & \matColumn{cyan};       \\ 
 \matColumn{red};   & \matColumn{red}; &  \matColumn{red};  &\matColumn{red};   &\matColumn{red};   &\matColumn{green};   &\matColumn{green};   &\matColumn{green};   &\matColumn{green};   & \matColumn{green};  & \matColumn{cyan};  & \matColumn{cyan};  & \matColumn{cyan};  & \matColumn{cyan};  & \matColumn{cyan};       \\ 
 \matColumn{red};   & \matColumn{red}; &  \matColumn{red};  &\matColumn{red};   &\matColumn{red};   &\matColumn{green};   &\matColumn{green};   &\matColumn{green};   &\matColumn{green};   & \matColumn{green};  & \matColumn{cyan};  & \matColumn{cyan};  & \matColumn{cyan};  & \matColumn{cyan};  & \matColumn{cyan};       \\ 
\matColumn{green};   & \matColumn{green}; & \matColumn{green}; & \matColumn{green}; & \matColumn{green}; & \matColumn{yellow}; & \matColumn{yellow};   & \matColumn{yellow};   & \matColumn{yellow};   & \matColumn{yellow};   &  \matColumn{violet};  &\matColumn{violet};   &\matColumn{violet};   &\matColumn{violet};   & \matColumn{violet};     \\ 
\matColumn{green};   & \matColumn{green}; & \matColumn{green}; & \matColumn{green}; & \matColumn{green}; & \matColumn{yellow}; & \matColumn{yellow};   & \matColumn{yellow};   & \matColumn{yellow};   & \matColumn{yellow};   &  \matColumn{violet};  &\matColumn{violet};   &\matColumn{violet};   &\matColumn{violet};   & \matColumn{violet};     \\ 
\matColumn{green};   & \matColumn{green}; & \matColumn{green}; & \matColumn{green}; & \matColumn{green}; & \matColumn{yellow}; & \matColumn{yellow};   & \matColumn{yellow};   & \matColumn{yellow};   & \matColumn{yellow};   &  \matColumn{violet};  &\matColumn{violet};   &\matColumn{violet};   &\matColumn{violet};   & \matColumn{violet};     \\ 
\matColumn{green};   & \matColumn{green}; & \matColumn{green}; & \matColumn{green}; & \matColumn{green}; & \matColumn{yellow}; & \matColumn{yellow};   & \matColumn{yellow};   & \matColumn{yellow};   & \matColumn{yellow};   &  \matColumn{violet};  &\matColumn{violet};   &\matColumn{violet};   &\matColumn{violet};   & \matColumn{violet};     \\ 
\matColumn{green};   & \matColumn{green}; & \matColumn{green}; & \matColumn{green}; & \matColumn{green}; & \matColumn{yellow}; & \matColumn{yellow};   & \matColumn{yellow};   & \matColumn{yellow};   & \matColumn{yellow};   &  \matColumn{violet};  &\matColumn{violet};   &\matColumn{violet};   &\matColumn{violet};   & \matColumn{violet};     \\ 
 \matColumn{cyan};    & \matColumn{cyan};  & \matColumn{cyan};  & \matColumn{cyan};  & \matColumn{cyan};  & \matColumn{violet}; &  \matColumn{violet};  &  \matColumn{violet}; &  \matColumn{violet};  &  \matColumn{violet}; & \matColumn{orange}; &\matColumn{orange};  &\matColumn{orange};  &\matColumn{orange};  &\matColumn{orange};      \\ 
 \matColumn{cyan};    & \matColumn{cyan};  & \matColumn{cyan};  & \matColumn{cyan};  & \matColumn{cyan};  & \matColumn{violet}; &  \matColumn{violet};  &  \matColumn{violet}; &  \matColumn{violet};  &  \matColumn{violet}; & \matColumn{orange}; &\matColumn{orange};  &\matColumn{orange};  &\matColumn{orange};  &\matColumn{orange};      \\ 
 \matColumn{cyan};    & \matColumn{cyan};  & \matColumn{cyan};  & \matColumn{cyan};  & \matColumn{cyan};  & \matColumn{violet}; &  \matColumn{violet};  &  \matColumn{violet}; &  \matColumn{violet};  &  \matColumn{violet}; & \matColumn{orange}; &\matColumn{orange};  &\matColumn{orange};  &\matColumn{orange};  &\matColumn{orange};      \\ 
 \matColumn{cyan};    & \matColumn{cyan};  & \matColumn{cyan};  & \matColumn{cyan};  & \matColumn{cyan};  & \matColumn{violet}; &  \matColumn{violet};  &  \matColumn{violet}; &  \matColumn{violet};  &  \matColumn{violet}; & \matColumn{orange}; &\matColumn{orange};  &\matColumn{orange};  &\matColumn{orange};  &\matColumn{orange};      \\ 
 \matColumn{cyan};    & \matColumn{cyan};  & \matColumn{cyan};  & \matColumn{cyan};  & \matColumn{cyan};  & \matColumn{violet}; &  \matColumn{violet};  &  \matColumn{violet}; &  \matColumn{violet};  &  \matColumn{violet}; & \matColumn{orange}; &\matColumn{orange};  &\matColumn{orange};  &\matColumn{orange};  &\matColumn{orange};      \\ 
  };
  \end{scope}
  \draw[->] (2,0) -- (3.5,0);
  \draw(2.75, 0.7) node[anchor=south]{Diagonalize};
  \draw(2.75, 0.4) node[anchor=south]{with};
  \draw(2.75, 0.0) node[anchor=south]{FFT};
  
  \begin{scope}[xshift=5.5cm]
     \matrix[mtn] (Y) {%
   \matDiag{red}; &  \matDiag{red}; &  \matDiag{red};& \matDiag{red}; &\matDiag{red};  &\matDiag{green};  &\matDiag{green};  &\matDiag{green};  &\matDiag{green};  & \matDiag{green};   &\matDiag{cyan};   & \matDiag{cyan};  &  \matDiag{cyan}; & \matDiag{cyan};  &  \matDiag{cyan};    \\ 
 \matDiag{red};   & \matDiag{red}; &  \matDiag{red};  &\matDiag{red};   &\matDiag{red};   &\matDiag{green};   &\matDiag{green};   &\matDiag{green};   &\matDiag{green};   & \matDiag{green};  & \matDiag{cyan};  & \matDiag{cyan};  & \matDiag{cyan};  & \matDiag{cyan};  & \matDiag{cyan};       \\ 
 \matDiag{red};   & \matDiag{red}; &  \matDiag{red};  &\matDiag{red};   &\matDiag{red};   &\matDiag{green};   &\matDiag{green};   &\matDiag{green};   &\matDiag{green};   & \matDiag{green};  & \matDiag{cyan};  & \matDiag{cyan};  & \matDiag{cyan};  & \matDiag{cyan};  & \matDiag{cyan};       \\ 
 \matDiag{red};   & \matDiag{red}; &  \matDiag{red};  &\matDiag{red};   &\matDiag{red};   &\matDiag{green};   &\matDiag{green};   &\matDiag{green};   &\matDiag{green};   & \matDiag{green};  & \matDiag{cyan};  & \matDiag{cyan};  & \matDiag{cyan};  & \matDiag{cyan};  & \matDiag{cyan};       \\ 
 \matDiag{red};   & \matDiag{red}; &  \matDiag{red};  &\matDiag{red};   &\matDiag{red};   &\matDiag{green};   &\matDiag{green};   &\matDiag{green};   &\matDiag{green};   & \matDiag{green};  & \matDiag{cyan};  & \matDiag{cyan};  & \matDiag{cyan};  & \matDiag{cyan};  & \matDiag{cyan};       \\ 
  \matDiag{green};   &  \matDiag{green};  &  \matDiag{green};  &  \matDiag{green};  &  \matDiag{green};  & \matDiag{yellow}; & \matDiag{yellow};   & \matDiag{yellow};   & \matDiag{yellow};   & \matDiag{yellow};   &  \matDiag{violet};  &\matDiag{violet};   &\matDiag{violet};   &\matDiag{violet};   & \matDiag{violet};     \\ 
    \matDiag{green};   &  \matDiag{green};  &  \matDiag{green};  &  \matDiag{green};  &  \matDiag{green};  & \matDiag{yellow}; & \matDiag{yellow};   & \matDiag{yellow};   & \matDiag{yellow};   & \matDiag{yellow};   &  \matDiag{violet};  &\matDiag{violet};   &\matDiag{violet};   &\matDiag{violet};   & \matDiag{violet};     \\    
      \matDiag{green};   &  \matDiag{green};  &  \matDiag{green};  &  \matDiag{green};  &  \matDiag{green};  & \matDiag{yellow}; & \matDiag{yellow};   & \matDiag{yellow};   & \matDiag{yellow};   & \matDiag{yellow};   &  \matDiag{violet};  &\matDiag{violet};   &\matDiag{violet};   &\matDiag{violet};   & \matDiag{violet};     \\ 
   \matDiag{green};   &  \matDiag{green};  &  \matDiag{green};  &  \matDiag{green};  &  \matDiag{green};  & \matDiag{yellow}; & \matDiag{yellow};   & \matDiag{yellow};   & \matDiag{yellow};   & \matDiag{yellow};   &  \matDiag{violet};  &\matDiag{violet};   &\matDiag{violet};   &\matDiag{violet};   & \matDiag{violet};     \\ 
  \matDiag{green};   &  \matDiag{green};  &  \matDiag{green};  &  \matDiag{green};  &  \matDiag{green};  & \matDiag{yellow}; & \matDiag{yellow};   & \matDiag{yellow};   & \matDiag{yellow};   & \matDiag{yellow};   &  \matDiag{violet};  &\matDiag{violet};   &\matDiag{violet};   &\matDiag{violet};   & \matDiag{violet};     \\ 
   \matDiag{cyan}; &  \matDiag{cyan};  &  \matDiag{cyan};  &  \matDiag{cyan};  & \matDiag{cyan};   &  \matDiag{violet};  &  \matDiag{violet};  &\matDiag{violet};  & \matDiag{violet}; & \matDiag{violet}; & \matDiag{orange}; &\matDiag{orange};  &\matDiag{orange};  &\matDiag{orange};  &\matDiag{orange};      \\ 
     \matDiag{cyan}; &  \matDiag{cyan};  &  \matDiag{cyan};  &  \matDiag{cyan};  & \matDiag{cyan};   &  \matDiag{violet};  &  \matDiag{violet};  &\matDiag{violet};  & \matDiag{violet}; & \matDiag{violet}; & \matDiag{orange}; &\matDiag{orange};  &\matDiag{orange};  &\matDiag{orange};  &\matDiag{orange};      \\ 
     \matDiag{cyan}; &  \matDiag{cyan};  &  \matDiag{cyan};  &  \matDiag{cyan};  & \matDiag{cyan};   &  \matDiag{violet};  &  \matDiag{violet};  &\matDiag{violet};  & \matDiag{violet}; & \matDiag{violet}; & \matDiag{orange}; &\matDiag{orange};  &\matDiag{orange};  &\matDiag{orange};  &\matDiag{orange};      \\ 
    \matDiag{cyan}; &  \matDiag{cyan};  &  \matDiag{cyan};  &  \matDiag{cyan};  & \matDiag{cyan};   &  \matDiag{violet};  &  \matDiag{violet};  &\matDiag{violet};  & \matDiag{violet}; & \matDiag{violet}; & \matDiag{orange}; &\matDiag{orange};  &\matDiag{orange};  &\matDiag{orange};  &\matDiag{orange};      \\ 
   \matDiag{cyan}; &  \matDiag{cyan};  &  \matDiag{cyan};  &  \matDiag{cyan};  & \matDiag{cyan};   &  \matDiag{violet};  &  \matDiag{violet};  &\matDiag{violet};  & \matDiag{violet}; & \matDiag{violet}; & \matDiag{orange}; &\matDiag{orange};  &\matDiag{orange};  &\matDiag{orange};  &\matDiag{orange};      \\ 
  };
  \end{scope}
  
  \draw[->] (7.25,0) -- (9,0);
  \draw(8.125, 0.1) node[anchor=south]{Permute};

  \draw(11,2.5) node[anchor=south]{Block-diagonal};
    \draw(11, 2) node[anchor=south]{with};
    \draw(11,1.5) node[anchor=south]{BTTB blocks};

  \begin{scope}[xshift=11cm]
  \matrix[mtn] (Y) {%
   \matCol{red}; & \matCol{red}; & \matCol{green};& \matCol{green}; & \matCol{cyan}; & \matCol{cyan};  & &  & & & && &  & & & & & & & & & \\
   \matCol{red}; & \matCol{red}; & \matCol{green};& \matCol{green}; & \matCol{cyan}; & \matCol{cyan};  & &  & & & && &  & & & & & & & & & \\
   \matCol{green}; & \matCol{green}; & \matCol{yellow};& \matCol{yellow}; & \matCol{violet}; & \matCol{violet};  & &  & & & & & & && &  & & & & & & \\
   \matCol{green}; & \matCol{green}; & \matCol{yellow};& \matCol{yellow}; & \matCol{violet}; & \matCol{violet};  & &  & & & && &  & & & & & & & & & \\
   \matCol{cyan}; & \matCol{cyan}; & \matCol{violet};& \matCol{violet}; & \matCol{orange}; & \matCol{orange};  & &  & & & & & &  & & & && & & & & \\
    \matCol{cyan}; & \matCol{cyan}; & \matCol{violet};& \matCol{violet}; & \matCol{orange}; & \matCol{orange};  & &  & & & & & & & & & & &  & & & &\\
  & & & & & & \matCol{red}; & \matCol{red}; & \matCol{green};& \matCol{green}; & \matCol{cyan}; & \matCol{cyan};  & &  & & & && &  & & & &  \\
  & & && & & \matCol{red}; & \matCol{red}; & \matCol{green};& \matCol{green}; & \matCol{cyan}; & \matCol{cyan};  & &  & & & & & &  & & & & \\
  & & & && & \matCol{green}; & \matCol{green}; & \matCol{yellow};& \matCol{yellow}; & \matCol{violet}; & \matCol{violet};  & &  & & & && & & & & & \\
  & & & && & \matCol{green}; & \matCol{green}; & \matCol{yellow};& \matCol{yellow}; & \matCol{violet}; & \matCol{violet};  & &  & & & && & & & & & \\
  & & && & & \matCol{cyan}; & \matCol{cyan}; & \matCol{violet};& \matCol{violet}; & \matCol{orange}; & \matCol{orange};   & &  & & & && & & & & & \\
  & & && & & \matCol{cyan}; & \matCol{cyan}; & \matCol{violet};& \matCol{violet}; & \matCol{orange}; & \matCol{orange};  & &  & & & && & & & & & \\
   & && &  & & & & & & & & \matCol{red}; & \matCol{red}; & \matCol{green};& \matCol{green}; & \matCol{cyan}; & \matCol{cyan};  & &  & & & & \\
  & && &  & & & & && & & \matCol{red}; & \matCol{red}; & \matCol{green};& \matCol{green}; & \matCol{cyan}; & \matCol{cyan};  & &  & & & & \\
  & && &  & & & & & && & \matCol{green}; & \matCol{green}; & \matCol{yellow};& \matCol{yellow}; & \matCol{violet}; & \matCol{violet};  & &  & & & & \\
  & & & &  & & & && && & \matCol{green}; & \matCol{green}; & \matCol{yellow};& \matCol{yellow}; & \matCol{violet}; & \matCol{violet};  & &  & & & & \\
  & & & &  & & & &&& & & \matCol{cyan}; & \matCol{cyan}; & \matCol{violet};& \matCol{violet}; & \matCol{orange}; & \matCol{orange};  & &  & & & &  \\
  & & && &  & & & && & & \matCol{cyan}; & \matCol{cyan}; & \matCol{violet};& \matCol{violet}; & \matCol{orange}; & \matCol{orange};  & &  & & & &\\
   & & &  & & & &&& &  & & & & & & & & \matCol{red}; & \matCol{red}; & \matCol{green};& \matCol{green}; & \matCol{cyan}; & \matCol{cyan};   \\
 & &  & & & && && &  & & & & && & & \matCol{red}; & \matCol{red}; & \matCol{green};& \matCol{green}; & \matCol{cyan}; & \matCol{cyan};  \\
  & &  & & & && && &  & & & & & && & \matCol{green}; & \matCol{green}; & \matCol{yellow};& \matCol{yellow}; & \matCol{violet}; & \matCol{violet};   \\
 & &  & & & & & & & &  & & & && && & \matCol{green}; & \matCol{green}; & \matCol{yellow};& \matCol{yellow}; & \matCol{violet}; & \matCol{violet};   \\
 & &  & & & & & & & &  & & & &&& & & \matCol{cyan}; & \matCol{cyan}; & \matCol{violet};& \matCol{violet}; & \matCol{orange}; & \matCol{orange};    \\
 & &  & & & & & & && &  & & & && & & \matCol{cyan}; & \matCol{cyan}; & \matCol{violet};& \matCol{violet}; & \matCol{orange}; & \matCol{orange};  \\
  };
  \end{scope}

\end{tikzpicture}%

\caption{\textbf{The 1-level circulant approximation.} Once the circulant-block approximation is constructed, the blocks are diagonalized with the FFT to create a diagonal-block matrix, then this is permuted to obtain a block-diagonal matrix to be inverted. In practice, it is not necessary to perform the first two steps and one can create the final block-diagonal matrix directly. Note that in our problem, where we have three levels of Toeplitz structure, the blocks in the final matrix are themselves BTTB with two levels of Toeplitz structure. This allows the process to be repeated for each of the blocks thus creating a further level of circulant approximation.}

\label{fig:1-level}
\end{figure*}

T.\ Chan's preconditioner was extended to Toeplitz-block matrices in \cite{chan1994circulant}. In our setting, the DDA matrix, $\mathbf{G}$, has $(3mn)^2$ Toeplitz blocks, each of size $l\times l$. Let us denote such a matrix $\textbf{T}_B$ (although we should keep in mind our DDA matrix). Then its circulant-block approximation, $\textbf{C}_B$, is obtained by calculating the circulant approximation to each Toeplitz block via (\ref{eqn:Chan}). These matrices are written as  
\[
\textbf{T}_B = 
\left( \begin{array}{cccc}
\text{T}_{1,1} & \text{T}_{1,2} & \ldots & \text{T}_{1,3mn} \\
\text{T}_{2,1} & \text{T}_{2,2} & \ldots & \text{T}_{2,3mn} \\
\vdots & \vdots  &            & \vdots   \\
\text{T}_{3mn,1} & \text{T}_{3mn,2} & \ldots & \text{T}_{3mn,3mn}
\end{array} \right), 
\]
and
\[
\textbf{C}_B  = 
\left( \begin{array}{cccc}
\text{C}(\text{T}_{1,1}) & \text{C}(\text{T}_{1,2}) & \ldots & \text{C}(\text{T}_{1,3mn}) \\
\text{C}(\text{T}_{2,1}) & \text{C}(\text{T}_{2,2}) & \ldots & \text{C}(\text{T}_{2,3mn}) \\
\vdots & \vdots  &            & \vdots   \\
\text{C}(\text{T}_{3mn,1}) & \text{C}(\text{T}_{3mn,2}) & \ldots & \text{C}(\text{T}_{3mn,3mn})
\end{array} \right),
\]
where $\text{C}(\text{T})$ denotes the Chan circulant approximation, defined by (\ref{eqn:Chan}), to $\text{T}$. Having constructed $\textbf{C}_B$, we then proceed to calculate its inverse via applications of the FFT. Each circulant block of $\textbf{C}_B$ has the representation $\text{C}(\text{T}_{ij}) = \text{F}^{-1}\Lambda_{ij}\text{F}$. Defining $\textbf{F} = \text{I}\otimes\text{F}$, we then have that
\begin{equation}
			\textbf{C}_B = [\text{C}(\text{T}_{ij})]_{i,j=1}^{3mn} = [\text{F}^{-1}\Lambda_{ij}\text{F}]_{i,j=1}^{3mn} = \textbf{F}^{-1}[\Lambda_{ij}]_{i,j=1}^{3mn}\textbf{F}.
\end{equation}
The matrix $[\Lambda_{ij}]_{i,j=1}^{3mn}$ is an $3lmn\times 3lmn$ diagonal-block matrix, where the diagonal blocks have size $l\times l$. As described in \cite{chan1994circulant}, this matrix is easily collapsed to a block-diagonal matrix $\textbf{D}$ after multiplication by a permutation matrix $\textbf{P}$, where
\begin{equation}
				\text{diag}(\text{D}_1,\ldots,\text{D}_l) = \textbf{P}[\Lambda_{ij}]_{i,j=1}^{3mn}\textbf{P}^{\text{T}}.
\end{equation}
Therefore, the inverse of $\textbf{C}_B$ is given by
\begin{equation}
				\textbf{C}_B^{-1} = \textbf{F}^{-1}\textbf{P}^{\text{T}}\text{diag}(\text{D}_1^{-1},\ldots,\text{D}_l^{-1})\textbf{P}\textbf{F}.
\end{equation}
We term $\textbf{C}_B$ the \textit{1-level circulant preconditioner} and illustrate its construction in Figure~\ref{fig:1-level}. The cost of the inversion of $\textbf{C}_B$ is dominated by the inversion of the $l$ dense blocks $\text{D}_i$, each of size $3mn\times 3mn$. Therefore, the cost is $\mathcal{O}(l(3nm)^3)$. 

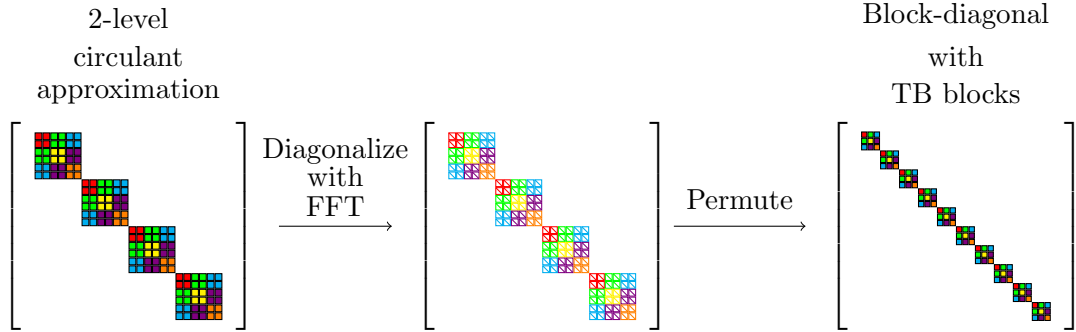
\begin{figure*}
\centering
\begin{tikzpicture}[
  node distance = 0.3cm,
  mtn/.style = {
    matrix of nodes,
    left delimiter = {[},
    right delimiter = {]}
  },
  ]
    \draw(0,2.5) node[anchor=south]{2-level};
    \draw(0, 2) node[anchor=south]{circulant};
    \draw(0,1.5) node[anchor=south]{approximation};
    \begin{scope}
\matrix[mtn] (Y) {%
   \matCol{red}; & \matCol{red}; & \matCol{green};& \matCol{green}; & \matCol{cyan}; & \matCol{cyan};  & &  & & & && &  & & & & & & & & & \\
   \matCol{red}; & \matCol{red}; & \matCol{green};& \matCol{green}; & \matCol{cyan}; & \matCol{cyan};  & &  & & & && &  & & & & & & & & & \\
   \matCol{green}; & \matCol{green}; & \matCol{yellow};& \matCol{yellow}; & \matCol{violet}; & \matCol{violet};  & &  & & & & & & && &  & & & & & & \\
   \matCol{green}; & \matCol{green}; & \matCol{yellow};& \matCol{yellow}; & \matCol{violet}; & \matCol{violet};  & &  & & & && &  & & & & & & & & & \\
   \matCol{cyan}; & \matCol{cyan}; & \matCol{violet};& \matCol{violet}; & \matCol{orange}; & \matCol{orange};  & &  & & & & & &  & & & && & & & & \\
    \matCol{cyan}; & \matCol{cyan}; & \matCol{violet};& \matCol{violet}; & \matCol{orange}; & \matCol{orange};  & &  & & & & & & & & & & &  & & & &\\
  & & & & & & \matCol{red}; & \matCol{red}; & \matCol{green};& \matCol{green}; & \matCol{cyan}; & \matCol{cyan};  & &  & & & && &  & & & &  \\
  & & && & & \matCol{red}; & \matCol{red}; & \matCol{green};& \matCol{green}; & \matCol{cyan}; & \matCol{cyan};  & &  & & & & & &  & & & & \\
  & & & && & \matCol{green}; & \matCol{green}; & \matCol{yellow};& \matCol{yellow}; & \matCol{violet}; & \matCol{violet};  & &  & & & && & & & & & \\
  & & & && & \matCol{green}; & \matCol{green}; & \matCol{yellow};& \matCol{yellow}; & \matCol{violet}; & \matCol{violet};  & &  & & & && & & & & & \\
  & & && & & \matCol{cyan}; & \matCol{cyan}; & \matCol{violet};& \matCol{violet}; & \matCol{orange}; & \matCol{orange};   & &  & & & && & & & & & \\
  & & && & & \matCol{cyan}; & \matCol{cyan}; & \matCol{violet};& \matCol{violet}; & \matCol{orange}; & \matCol{orange};  & &  & & & && & & & & & \\
   & && &  & & & & & & & & \matCol{red}; & \matCol{red}; & \matCol{green};& \matCol{green}; & \matCol{cyan}; & \matCol{cyan};  & &  & & & & \\
  & && &  & & & & && & & \matCol{red}; & \matCol{red}; & \matCol{green};& \matCol{green}; & \matCol{cyan}; & \matCol{cyan};  & &  & & & & \\
  & && &  & & & & & && & \matCol{green}; & \matCol{green}; & \matCol{yellow};& \matCol{yellow}; & \matCol{violet}; & \matCol{violet};  & &  & & & & \\
  & & & &  & & & && && & \matCol{green}; & \matCol{green}; & \matCol{yellow};& \matCol{yellow}; & \matCol{violet}; & \matCol{violet};  & &  & & & & \\
  & & & &  & & & &&& & & \matCol{cyan}; & \matCol{cyan}; & \matCol{violet};& \matCol{violet}; & \matCol{orange}; & \matCol{orange};  & &  & & & &  \\
  & & && &  & & & && & & \matCol{cyan}; & \matCol{cyan}; & \matCol{violet};& \matCol{violet}; & \matCol{orange}; & \matCol{orange};  & &  & & & &\\
   & & &  & & & &&& &  & & & & & & & & \matCol{red}; & \matCol{red}; & \matCol{green};& \matCol{green}; & \matCol{cyan}; & \matCol{cyan};   \\
 & &  & & & && && &  & & & & && & & \matCol{red}; & \matCol{red}; & \matCol{green};& \matCol{green}; & \matCol{cyan}; & \matCol{cyan};  \\
  & &  & & & && && &  & & & & & && & \matCol{green}; & \matCol{green}; & \matCol{yellow};& \matCol{yellow}; & \matCol{violet}; & \matCol{violet};   \\
 & &  & & & & & & & &  & & & && && & \matCol{green}; & \matCol{green}; & \matCol{yellow};& \matCol{yellow}; & \matCol{violet}; & \matCol{violet};   \\
 & &  & & & & & & & &  & & & &&& & & \matCol{cyan}; & \matCol{cyan}; & \matCol{violet};& \matCol{violet}; & \matCol{orange}; & \matCol{orange};    \\
 & &  & & & & & & && &  & & & && & & \matCol{cyan}; & \matCol{cyan}; & \matCol{violet};& \matCol{violet}; & \matCol{orange}; & \matCol{orange};  \\
  };

  \end{scope}

  \draw[->] (2,0) -- (3.5,0);
  \draw(2.75, 0.7) node[anchor=south]{Diagonalize};
  \draw(2.75, 0.4) node[anchor=south]{with};
  \draw(2.75, 0.0) node[anchor=south]{FFT};

  \begin{scope}[xshift=5.5cm]
  \matrix[mtn] (Y) {%
   \matdiag{red}; & \matdiag{red}; & \matdiag{green};& \matdiag{green}; & \matdiag{cyan}; & \matdiag{cyan};  & &  & & & && &  & & & & & & & & & \\
   \matdiag{red}; & \matdiag{red}; & \matdiag{green};& \matdiag{green}; & \matdiag{cyan}; & \matdiag{cyan};  & &  & & & && &  & & & & & & & & & \\
   \matdiag{green}; & \matdiag{green}; & \matdiag{yellow};& \matdiag{yellow}; & \matdiag{violet}; & \matdiag{violet};  & &  & & & & & & && &  & & & & & & \\
   \matdiag{green}; & \matdiag{green}; & \matdiag{yellow};& \matdiag{yellow}; & \matdiag{violet}; & \matdiag{violet};  & &  & & & && &  & & & & & & & & & \\
   \matdiag{cyan}; & \matdiag{cyan}; & \matdiag{violet};& \matdiag{violet}; & \matdiag{orange}; & \matdiag{orange};  & &  & & & & & &  & & & && & & & & \\
    \matdiag{cyan}; & \matdiag{cyan}; & \matdiag{violet};& \matdiag{violet}; & \matdiag{orange}; & \matdiag{orange};  & &  & & & & & & & & & & &  & & & &\\
  & & & & & & \matdiag{red}; & \matdiag{red}; & \matdiag{green};& \matdiag{green}; & \matdiag{cyan}; & \matdiag{cyan};  & &  & & & && &  & & & &  \\
  & & && & & \matdiag{red}; & \matdiag{red}; & \matdiag{green};& \matdiag{green}; & \matdiag{cyan}; & \matdiag{cyan};  & &  & & & & & &  & & & & \\
  & & & && & \matdiag{green}; & \matdiag{green}; & \matdiag{yellow};& \matdiag{yellow}; & \matdiag{violet}; & \matdiag{violet};  & &  & & & && & & & & & \\
  & & & && & \matdiag{green}; & \matdiag{green}; & \matdiag{yellow};& \matdiag{yellow}; & \matdiag{violet}; & \matdiag{violet};  & &  & & & && & & & & & \\
  & & && & & \matdiag{cyan}; & \matdiag{cyan}; & \matdiag{violet};& \matdiag{violet}; & \matdiag{orange}; & \matdiag{orange};   & &  & & & && & & & & & \\
  & & && & & \matdiag{cyan}; & \matdiag{cyan}; & \matdiag{violet};& \matdiag{violet}; & \matdiag{orange}; & \matdiag{orange};  & &  & & & && & & & & & \\
   & && &  & & & & & & & & \matdiag{red}; & \matdiag{red}; & \matdiag{green};& \matdiag{green}; & \matdiag{cyan}; & \matdiag{cyan};  & &  & & & & \\
  & && &  & & & & && & & \matdiag{red}; & \matdiag{red}; & \matdiag{green};& \matdiag{green}; & \matdiag{cyan}; & \matdiag{cyan};  & &  & & & & \\
  & && &  & & & & & && & \matdiag{green}; & \matdiag{green}; & \matdiag{yellow};& \matdiag{yellow}; & \matdiag{violet}; & \matdiag{violet};  & &  & & & & \\
  & & & &  & & & && && & \matdiag{green}; & \matdiag{green}; & \matdiag{yellow};& \matdiag{yellow}; & \matdiag{violet}; & \matdiag{violet};  & &  & & & & \\
  & & & &  & & & &&& & & \matdiag{cyan}; & \matdiag{cyan}; & \matdiag{violet};& \matdiag{violet}; & \matdiag{orange}; & \matdiag{orange};  & &  & & & &  \\
  & & && &  & & & && & & \matdiag{cyan}; & \matdiag{cyan}; & \matdiag{violet};& \matdiag{violet}; & \matdiag{orange}; & \matdiag{orange};  & &  & & & &\\
   & & &  & & & &&& &  & & & & & & & & \matdiag{red}; & \matdiag{red}; & \matdiag{green};& \matdiag{green}; & \matdiag{cyan}; & \matdiag{cyan};   \\
 & &  & & & && && &  & & & & && & & \matdiag{red}; & \matdiag{red}; & \matdiag{green};& \matdiag{green}; & \matdiag{cyan}; & \matdiag{cyan};  \\
  & &  & & & && && &  & & & & & && & \matdiag{green}; & \matdiag{green}; & \matdiag{yellow};& \matdiag{yellow}; & \matdiag{violet}; & \matdiag{violet};   \\
 & &  & & & & & & & &  & & & && && & \matdiag{green}; & \matdiag{green}; & \matdiag{yellow};& \matdiag{yellow}; & \matdiag{violet}; & \matdiag{violet};   \\
 & &  & & & & & & & &  & & & &&& & & \matdiag{cyan}; & \matdiag{cyan}; & \matdiag{violet};& \matdiag{violet}; & \matdiag{orange}; & \matdiag{orange};    \\
 & &  & & & & & & && &  & & & && & & \matdiag{cyan}; & \matdiag{cyan}; & \matdiag{violet};& \matdiag{violet}; & \matdiag{orange}; & \matdiag{orange};  \\
  };  \end{scope}
  
  \draw[->] (7.25,0) -- (9,0);
  \draw(8.125, 0.1) node[anchor=south]{Permute};
  
    \draw(11,2.5) node[anchor=south]{Block-diagonal};
    \draw(11, 2) node[anchor=south]{with};
    \draw(11,1.5) node[anchor=south]{TB blocks};
  
   \begin{scope}[xshift=11cm]
  \matrix[mtn] (Y) {%
   \matcol{red}; & \matcol{green} & \matcol{cyan} & & & & & && & & & &&& & & && & & & & & && & && \\
   \matcol{green} & \matcol{yellow} & \matcol{violet} & & & & && & &&& & &  & & &&& & & & & && & && &\\
   \matcol{cyan} & \matcol{violet} & \matcol{orange} & & & && & && &  & & && & & & & &&& & &&& & & &\\
  & & & \matcol{red}; & \matcol{green} & \matcol{cyan}  & & & &  & & &&&& & && &&& & & & & & & &&\\
  & & &\matcol{green} & \matcol{yellow} & \matcol{violet}   & & && & && & & & & & && && && && && & &\\
  & & & \matcol{cyan} & \matcol{violet} & \matcol{orange} & && & && & & & & & & &&&& & & & & && & &\\
  & & && & & \matcol{red}; & \matcol{green} & \matcol{cyan} & & && && & & & && & &  & & && &&& & \\
  & & && & &\matcol{green} & \matcol{yellow} & \matcol{violet} & & && & && & & & & & & && & & &&& & \\
  & & && & & \matcol{cyan} & \matcol{violet} & \matcol{orange} & & && &  & & && & & && & & & &&&& &\\
   & & && & && & & \matcol{red}; & \matcol{green} & \matcol{cyan}  & & && & & && & &&& && & & & &  \\
  & & && & && & &\matcol{green} & \matcol{yellow} & \matcol{violet} & & && && & & & & & && & & && & \\
  & & && & && & & \matcol{cyan} & \matcol{violet} & \matcol{orange} && & & & && & &  & & &&& & & &&\\
     & & && & && & && & & \matcol{red}; & \matcol{green} & \matcol{cyan}  & & &  & & && & & & &&&& &\\
  & & && & &&& & & & &\matcol{green} & \matcol{yellow} & \matcol{violet}  & & && & && & & & && && & \\
  & & && & && & && & & \matcol{cyan} & \matcol{violet} & \matcol{orange}& & & & & & & & &&&  & & &&\\
       & & && & && & && & && & & \matcol{red}; & \matcol{green} & \matcol{cyan} && & & & & & & && & & \\
  & & && & &&&& & & & & & &\matcol{green} & \matcol{yellow} & \matcol{violet}  && & & & & & & && & &\\
  & & && & && & && && & & & \matcol{cyan} & \matcol{violet} & \matcol{orange} & &  & & &&& & & && &\\
       & & && & && & && & && & && & & \matcol{red}; & \matcol{green} & \matcol{cyan}&  & & && & && & \\
  & & && && & & &&&& & & & & & &\matcol{green} & \matcol{yellow} & \matcol{violet} &  & & && & && & \\
  & & && & && && & & && && & & & \matcol{cyan} & \matcol{violet} & \matcol{orange} && & & & & && &\\
    & & && & && & && & && && & & && & & \matcol{red}; & \matcol{green} & \matcol{cyan} & & & & & & \\
  & & && && & & &&&& & & && & & & & &\matcol{green} & \matcol{yellow} & \matcol{violet} & & & & & & \\
  & & && & && && & & && & && && & & & \matcol{cyan} & \matcol{violet} & \matcol{orange}  & & && & &\\
      & & && && & & && & && & && && & & && & & \matcol{red}; & \matcol{green} & \matcol{cyan}  & & & \\
  & & && & && && & & &&&& & & && & & & & &\matcol{green} & \matcol{yellow} & \matcol{violet}  & & & \\
  & & && & && & && && & & && & && && & & & \matcol{cyan} & \matcol{violet} & \matcol{orange} & & & \\
        & & & & & && && & & && & && & && && & & && & & \matcol{red}; & \matcol{green} & \matcol{cyan}  \\
  & & && & & & & && && & & &&&& & & && & & & & &\matcol{green} & \matcol{yellow} & \matcol{violet}  \\
  & & && & && &  & & &&& && & & && & && && & & & \matcol{cyan} & \matcol{violet} & \matcol{orange} \\
  };
  \end{scope}

  \end{tikzpicture}
  
  \caption{\textbf{The 2-level circulant approximation.} Once the circulant approximation to the blocks of the final matrix in Fig.\ref{fig:1-level} is constructed, the blocks are diagonalized with the FFT to create a diagonal-block matrix, then this is permuted to obtain a block-diagonal matrix to be inverted. Note that the blocks in the final matrix possess are Toeplitz, i.e., there is now only a single level of Toeplitz structure remaining. The process could therefore be repeated once again, however we do not consider 3-level circulant approximations in this paper.}
  
  \label{fig:2-level}
  
\end{figure*}

If $m$ and $n$ are small, $\textbf{C}_B$ can be a cheap preconditioner. If they are not small, one may resort to a second level of circulant approximation, applied this time to each of the dense blocks $\text{D}_i$. In our BTTB case, the blocks $\text{D}_i$ are themselves Toeplitz-block, thus allowing the above procedure to be repeated for each $\text{D}_i$, leading to a \textit{2-level circulant preconditioner} which we denote by $\textbf{C}_{B_2}$. The matrix $\textbf{C}_{B_2}$ can be written as
\begin{equation}
	\textbf{C}_{B_2} = \textbf{F}^{-1}\textbf{P}^{\text{T}}\text{diag}(\textbf{C}_{B_1}(\text{D}_1),\ldots,\textbf{C}_{B_1}(\text{D}_l))\textbf{P}\textbf{F},
\end{equation}
where $\textbf{C}_{B_1}(\text{D}_i)$ denotes the 1-level circulant approximation
\begin{equation}
	\textbf{C}_{B_1}(D_i) = \overline{\textbf{F}}^{-1}\overline{\textbf{P}}^{\text{T}}\text{diag}(\overline{\text{D}}_1,\ldots,\overline{\text{D}}_m)\overline{\textbf{P}}\overline{\textbf{F}},
\end{equation}
where $\overline{\text{D}}_i$ are new blocks, of size $3n\times 3n$. The lines above $\overline{\textbf{F}}$ and $\overline{\textbf{P}}$ are to highlight that they are of the dimension appropriate for $\overline{\text{D}}_i$. An illustration of the 2-level circulant approximation is shown in Figure~\ref{fig:2-level}. 
The resulting block-diagonal matrix has $lm$ blocks, each of size $3n\times 3n$, which must inverted to obtain $\textbf{C}_{B_2}^{-1}$ for use as our preconditioner:
\begin{equation}
	\textbf{C}^{-1}_{B_2} = \textbf{F}^{-1}\textbf{P}^{\text{T}}\text{diag}(\textbf{C}^{-1}_{B_1}(\text{D}_1),\ldots,\textbf{C}^{-1}_{B_1}(\text{D}_l))\textbf{P}\textbf{F}.
\end{equation}
Again, the inversion cost is dominated by the inversion of the $lm$ blocks and so is now $\mathcal{O}(lm(3n)^3)$.

\subsection{Algorithms}
\label{ss:algorithms}
We present a few details as to the practical construction and inversion of the 2-level circulant preconditioner $\textbf{C}_{B_2}$. First we remind the reader that matrix we are wishing to approximate is of the form
\[
	\boldsymbol{\alpha}^{-1} - \Delta^3\textbf{G}
\]
where $\boldsymbol{\alpha}$ is a diagonal matrix with each entry being the polarizability of the appropriate voxel and $\textbf{G}$ is our BTTB DDA matrix. In general, $\boldsymbol{\alpha}$ does not have a constant diagonal unless we are dealing with a homogeneous cuboid. In the examples considered in Section~\ref{sec:num_results}, we deal with homogeneous hexagonal prisms so that the polarizabilites take one of two values, that of the ``ice'' voxels or of the ``air'' voxels. 

The endeavour is to create the 2-level circulant approximation $\textbf{C}_{B_2}$ with the hope that $\textbf{C}_{B_2}\approx\boldsymbol{\alpha}^{-1} - \Delta^3\textbf{G}$ so that it acts as a good preconditioner. To do this, we first construct the 2-level circulant approximation to the BTTB matrix $\textbf{G}$ and denote this $\tilde{\textbf{C}}_{B_2}$. Then we must construct a constant diagonal matrix that approximates $\boldsymbol{\alpha}$, denoting this $\tilde{\alpha}\textbf{I}$. Now we have that the matrix
\begin{equation}
	\textbf{C}_{B_2} := \tilde{\alpha}^{-1}\textbf{I} - \Delta^3\tilde{\textbf{C}}_{B_2}
\end{equation}
is also circulant and hence appropriate as our circulant preconditioner. We stress the importance of the construction of the diagonal matrix $\tilde{\alpha}\textbf{I}$ since, when $\boldsymbol{\alpha}$ is not a constant diagonal (i.e., when the scatterer is not a homogeneous cuboid), the matrix $\boldsymbol{\alpha}^{-1} - \Delta^3\textbf{C}_{B_2}$ does not inherit the circulant properties of $\textbf{C}_{B_2}$ and hence is not cheaply inverted. The choice made here is to take $\tilde{\alpha}$ as the value of the ``ice'' voxels, however it may be the case that some $\tilde{\alpha}$ derived from averaging over the $\alpha_i$ leads to superior performance (as was seen in \cite{groth2019circulant} for the 1-level circulant preconditoner). But we do not explore this question in this article. The final step is the inversion of $\textbf{C}_{B_2}$, which is performed in a parallel loop over its $lm$ diagonal blocks, each of dimension $3n$.

\begin{table}[ht!]
	\centering
	\begin{tabular}{c |  c  c  c}
	                              & $1^{\text{st}}$ level ($x$) & $2^{\text{nd}}$ level ($y$) & $3^{\text{rd}}$ level ($z$) \\
	                              \hline
	$\mathbf{G}_{xx}$ & +  &  +   &    +  \\
	$\mathbf{G}_{xy}$ & --  &  --  &  + \\
	$\mathbf{G}_{xz}$ & -- & + & -- \\
	$\mathbf{G}_{yy}$ & + & + & + \\
	$\mathbf{G}_{yz}$ & + & -- & + \\
	$\mathbf{G}_{zz}$ & + & + & + \\
	\end{tabular}
	\caption{Symmetry (+) and anti-symmetry (--) patterns within the different levels of the BTTB structure of the constituent blocks of the matrix $\mathbf{G}$. This knowledge is required in the efficient construction of the circulant preconditioner.}
	\label{tab:symmetry}
\end{table}

Now we state a few details about the efficient construction of the circulant approximation to $\textbf{G}$. This construction can be performed efficiently by exploiting the symmetries within the constituent blocks of $\mathbf{G}$. Each of the six unique blocks of $\mathbf{G}$ has a three-level Toeplitz structure and on each of these levels the elements/blocks are arranged either symmetrically or anti-symmetrically - these symmetries are provided in Table~\ref{tab:symmetry}. 

As an example, let us take the block $\mathbf{G}_{xz}$. First we wish to calculate the 1-level circulant approximation to this block in the $x$-direction. We know that this block is anti-symmetric on the first level, so it's circulant approximation is given as shown in Algorithm~\ref{alg:Gxz}; note the minus sign in this version of the circulant approximation (\ref{eqn:Chan}). 
\begin{algorithm}
\caption{Compute 1-level circulant approximation to $\textbf{G}_{xz}$}
\label{array-sum}
\begin{algorithmic}[1]
\State $\textbf{G}_{xz} = \textbf{G}(:,:,:,3)$
\State $\mathbf{c}(1,:,:,) = \textbf{G}_{xz}(1,:,:)\quad$ the diagonal entries of the circulant and the Toeplitz matrices equal
\For {$i=1:l$}
	\State $\mathbf{c}(i,:,:) = -\frac{l+1-i}{l}\textbf{G}_{xz}(i,:,:) + \frac{i-1}{l}\textbf{G}_{xz}(l-i+2,:,:)\quad$ note the minus sign due to antisymmetry
\EndFor
\State $\tilde{\textbf{C}}^{(1)}_{xz} = \text{FFT}(\mathbf{c})\quad$ provides the Fourier transform of columns of $\mathbf{c}$
\end{algorithmic}
\label{alg:Gxz}
\end{algorithm}
After having performed this circulant approximation for each of the six unique portions of $\textbf{G}$ (taking into account the symmetry or anti-symmetry of each), we generate the defining tensor $\tilde{\textbf{C}}^{(1)}$ of the 1-level circulant preconditioner. From this, the full 1-level circulant, as shown in Figure~\ref{fig:1-level}, can be constructed. 

Here, however, we focus on the 2-level preconditioner so perform one further level of circulant approximation. Considering again the block $\textbf{G}_{xz}$, we must perform the circulant approximation in the $y$-direction to the tensor $\tilde{\textbf{C}}^{(1)}$, constructed in Algorithm~\ref{alg:Gxz}. This is shown in Algorithm~\ref{alg:Gxz_2}. Observe how we now loop over the $l$ blocks generated from the first level of circulant approximation. From the tensor $\tilde{\textbf{C}}^{(2)}$ we may now generate the full 2-level circulant approximation $\tilde{\textbf{C}}_{B_2}$ as shown in Figure~\ref{fig:2-level}. This generation of the full preconditioner requires some familiarity with the symmetries of the matrix $\textbf{G}$ as described in Table~\ref{tab:symmetry} but is not too complicated.
\begin{algorithm}
\caption{Compute 2-level circulant approximation to $\textbf{G}_{xz}$}
\label{array-sum}
\begin{algorithmic}[1]
\For {$i=1:l$}
	\State $\tilde{\textbf{C}}^{(1)}_{xz}(i,:,:) = \tilde{\textbf{C}}^{(1)}(i,:,:,3)$
	\State $\mathbf{c}(1,:,) = \tilde{\textbf{C}}^{(1)}_{xz}(i,1,:)\quad$ the diagonal entries of the circulant and the Toeplitz matrices equal
	\For {$j=1:m$}
		\State $\mathbf{c}(j,:) = \frac{m+1-i}{m}\tilde{\textbf{C}}^{(1)}_{xz}(i,j,:) + \frac{i-1}{l}\tilde{\textbf{C}}^{(1)}_{xz}(i,m-j+2,:)$
	\EndFor
	\State $\tilde{\textbf{C}}^{(2)}_{xz}(i,:,:) = \text{FFT}(\mathbf{c})\quad$ provides the Fourier transform of columns of $\mathbf{c}$
\EndFor
\end{algorithmic}
\label{alg:Gxz_2}
\end{algorithm}

Algorithm~\ref{alg:Gxz} and Algorithm~\ref{alg:Gxz_2} serve to illustrate that the generation of the 1- and 2-level circulant approximations are fairly straightforward and can be performed directly from the defining tensor $\textbf{G}\in\mathbb{C}^{l\ \times\ m\ \times\ n\ \times\ 6}$ given in Algorithm~\ref{alg:G}, which is  constructed within all FFT-accelerated DDA implementations.

\subsection{Costings}
\label{subsec:costings}
Costings were provided in Chan and Olkin~\cite{chan1994circulant} but for a general Toeplitz-block matrix. Here we provide the relevant costings for our symmetric system. Following~\cite{chan1994circulant} we consider a floating-point operation (flop) as one multiplication plus one addition. We also denote the cost of applying the FFT to a vector of length $n$ as fft($n$), which is typically $5\log_2(n)$ flops for standard FFT algorithms such as FFTW~\cite{frigo2005design}. 

First we consider the setup (including the inversion) and per-iteration application costs of the 1-level circulant preconditioner.

\textbf{1-level: setup}
\begin{enumerate}
	\item Point-circulant approximation via (\ref{eqn:Chan}) of the $6mn$ unique blocks of length~$l$: $6lmn$ flops 
	\item $6mn$ FFTs of length $l$ to generate $\tilde{\textbf{C}}^{(1)}$: $6mn\cdot\text{fft}(l)$
	\item Generate $l$ diagonal blocks from $\tilde{\textbf{C}}^{(1)}$ using knowledge of BTTB structure and symmetries in Table~\ref{tab:symmetry}: $\sim$ free.
	\item Inversion of the $l$ dense diagonal blocks of size $(3mn)\times(3mn)$: $\frac{1}{3}l(3mn)^3$.
\end{enumerate}
\[
	\text{1-level setup cost} = \frac{1}{3}l(3mn)^3 + 6lmn + 6mn\cdot\text{fft}(l).
\]

\textbf{1-level: application (per iteration)}
\begin{enumerate}
	\item $3mn$ FFTs, each of length $l$: $3mn\cdot\text{fft}(l)$
	\item Multiplication with the $l$ diagonal blocks $\text{D}_i$ of size $(3mn)\times(3mn)$: $l(3mn)^2$.
	\item $3mn$ inverse FFTs, each of length $l$: $3mn\cdot\text{fft}(l)$
\end{enumerate}
\[
	\text{1-level application cost} = l(3mn)^2 + 6mn\cdot\text{fft}(l).
\]

Next we consider the setup and per-iteration application costs of the 2-level circulant preconditioner.

\textbf{2-level setup}
\begin{enumerate}
	\item Steps 1 and 2 from the 1-level setup to generate the $l$ diagonal blocks: $6lmn + 6mn\cdot\text{fft}(l)$
	\item Point-circulant approximation via (\ref{eqn:Chan}) of the $6ln$ blocks of length $m$: $6lmn$.
	\item $l\times 6n$ FFTs of length $m$ to generate $\tilde{\textbf{C}}^{(1)}$: $6ln\cdot\text{fft}(m)$
	\item Generate $lm$ diagonal blocks from $\tilde{\textbf{C}}^{(2)}$ using knowledge of BTTB structure and symmetries in Table~\ref{tab:symmetry}: $\sim$ free.
	\item Inversion of each of the $lm$ dense diagonal blocks of size $(3n)\times(3n)$: $\frac{1}{3}lm(n)^3$.
\end{enumerate}
\begin{align*}
	\text{2-level setup cost} = & \frac{1}{3}lm(3n)^3 + l(6mn+6n\cdot\text{fft}(m)) \\
	&+ 6lmn + 6mn\cdot\text{fft}(l).
\end{align*}

\textbf{2-level application (per iteration)}
\begin{enumerate}
	\item $3mn$ FFTs of length $l$: $3mn\cdot\text{fft}(l)$
	\item For each of the $l$ blocks: 
		\begin{enumerate}[(i)]
			\item $3n$ FFTs of length $m$: $3n\cdot\text{fft}(m)$
			\item Multiplication with the $m$ blocks of size $(3n)\times(3n)$: $m(3n)^2$
			\item $3n$ inverse FFTs of length $m$: $3n\cdot\text{fft}(m)$
		\end{enumerate}
	\item $3mn$ inverse FFTs of length $l$: $3mn\cdot\text{fft}(l)$.
\end{enumerate}
\[
		\text{2-level application cost} = lm(3n)^2 + 6mn\cdot\text{fft}(l) + 6ln\cdot\text{fft}(m).
\]

The setup cost of the 1-level preconditioner is dominated by the block inversion and so the complexity is $\mathcal{O}(lm^3n^3)$. For problems in which $m,n\ll l$, this cost is low and hence this preconditioner is feasible - this was seen for silicon photonics geometries in \cite{groth2019circulant}. However, for ice crystal applications the geometries of interest are typically optically large in all three dimensions. For example, a cube with ten wavelengths across (a size parameter of $\sim 31$) discretized at a resolution of 10 voxels per wavelength would require the storage and inversion of 100 dense matrix blocks of dimension $3\times 10^4$, a cost that is extremely demanding of most computers. Switching instead to the 2-level preconditioner, for this example, requires the storage and inversion of $10^4$ dense matrix blocks of dimension $3\times 10^2$, a much more manageable task. Furthermore, this task can be performed in parallel and hence very rapidly, as we shall see in Section~\ref{sec:num_results}. For this reason of cost, we shall be applying only the 2-level preconditioner in this paper. For details on the performance of the 1-level, the reader is referred to \cite{groth2019circulant}. We also note that a 3-level circulant approximation is possible, however it was found in a preliminary study to yield poorer results so we do not consider it here.

\begin{table}[ht!]
	\centering
	\begin{tabular}{c | r | l}
		\multirow{3}{*}{\textbf{Preconditioner}} & Storage cost (memory) & $\mathcal{O}(x^4)$ \\
				       			       & Setup cost (time)	      & $\mathcal{O}(x^5)$ \\
							       & MVP (time)                    & $\mathcal{O}(x^4)$ \\
		\hline\hline
		\multirow{3}{*}{\textbf{Operator}}	       & Storage cost (memory) & $\mathcal{O}(x^3)$ \\
				       			       & Setup cost (time)	      & $\mathcal{O}(x^3)$ \\
							       & MVP (time)                    & $\mathcal{O}(x^3\log_2 x)$ \\		       
	\end{tabular}
	\caption{Costings for the preconditioner and integral operator in terms of size parameter. These are derived from expressions given in the text body by assuming $l,m,n\sim x$.}
	\label{tab:costings}
\end{table}
In terms of size parameter $x$, since $l,m,n\sim x$, we summarize the costings in Table~\ref{tab:costings}. For reference, we also provide the costings for assembling the defining tensor of $\textbf{G}$ and performing an MVP with it. We observe from the table that the cost of the preconditioner is greater than that of the integral operator, however not substantially so. Furthermore, the constants are hidden. In Section~\ref{sec:num_results}, we provide timings and memory consumption figures in order to observe these costs in practice.

\section{Numerical results}
\label{sec:num_results}
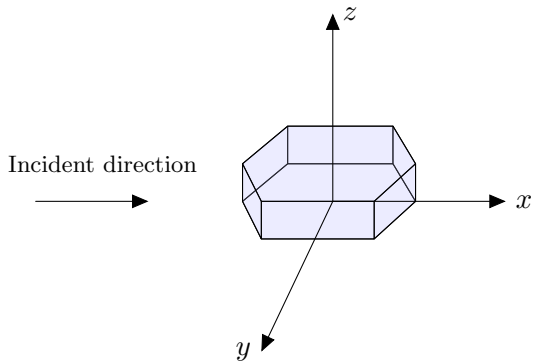
\begin{figure}[!h]
\centering
\begin{tikzpicture}[line cap=round,line join=round,>=triangle 45,x=1.0cm,y=1.0cm, scale=1]
    \tikzstyle{conefill} = [fill=blue!20,fill opacity=0.2]
    \tikzstyle{ann} = [fill=white,font=\footnotesize,inner sep=1pt]
    \tikzstyle{ghostfill} = [fill=white]
         \tikzstyle{ghostdraw} = [draw=black!50]
    \filldraw[conefill](-0.75,1.5)--(0.75,1.5)--(1.3,2)
                        --(1,2.5)--(-0.4,2.5)--(-1,2)--cycle;
    \filldraw[conefill](-1,2)--(-0.4,2.5)--(-0.4,6.5-3.5)--(-1,6-3.5)--cycle;
    \filldraw[conefill](-0.4,2.5)--(1,2.5)--(1,6.5-3.5)--(-0.4,6.5-3.5)--cycle;
    \filldraw[conefill](1,2.5)--(1,6.5-3.5)--(1.3,6-3.5)--(1.3,2)--cycle;
    \filldraw[conefill](1.3,2)--(1.3,6-3.5)--(0.75,5.5-3.5)--(0.75,1.5)--cycle;
    \filldraw[conefill](0.75,1.5)--(0.75,5.5-3.5)--(-0.75,5.5-3.5)--(-0.75,1.5)--cycle;
    \filldraw[conefill](-0.75,1.5)--(-0.75,5.5-3.5)--(-1,6-3.5)--(-1,2)--cycle;
    \filldraw[conefill](-0.75,5.5-3.5)--(0.75,5.5-3.5)--(1.3,6-3.5)
                        --(1,6.5-3.5)--(-0.4,6.5-3.5)--(-1,6-3.5)--cycle;
    
    \draw[->](0.2,2)--(2.5,2)node[right]{$x$};
    \draw[->](0.2,2)--(0.2,4.5)node[right]{$z$};
    \draw[->](0.2,2)--(-0.75,0)node[left]{$y$};
    
    \draw[->] (-3.75,2)--(-2.25,2);
    \draw (-4.25,2.5)node[right]{\footnotesize{Incident direction}};
    
\end{tikzpicture}
\caption{Geometrical setup: a $z$-polarized plane wave travelling in the positive $x$-direction is incident upon a hexagonal plate of aspect ratio $L/a=0.1, 0.2$.}
\label{fig:planes}
\end{figure}

In order to test the performance of the circulant preconditioner for a realistic application, we consider the scattering of a plane wave by hexagonal plates with a variety of refractive indices and size parameters. In particular, we consider the scattering setup shown in Figure~\ref{fig:planes}. 
The incident wave is polarized in the $z$-direction and travels in the positive $x$-direction, i.e., it has the form $\textbf{E}^{\text{inc}} = (0, 0, 1)e^{ikx}$. We consider two different values for the aspect ratio~$L/a = 0.1,\ 0.2$, where $L$ is the height of the plate and $a$ is the radius of the smallest circumscribing circle of the hexagonal face. The refractive indices considered are $\mu=1.2, 1.4, 1.6, 1.8, 2$ and the size parameters are $x=10, 20, 30, 40, 60, 80, 100$, where $x=ka$. These parameter values are chosen to allow for a soft comparison to \cite{yurkin2007discrete} where iteration counts are given for DDA, albeit there for scattering by spheres. 

We present performance results for the iterative solves of the linear system using both the generalized minimum residual method (GMRES) and the biconjugate gradient stabilized method (BiCG-Stab) on an Intel (R) Xeon (R) CPU E5-2680 v4 @ 2.40GHz machine. BiCG-Stab is a popular iterative solver for DDA since it is fast, however its convergence is not guaranteed. GMRES on the other hand is slower and more memory intensive owing to the storage and use of the Krylov vectors, but its convergence is guaranteed if the entire Krylov subspace is kept, which we do here. We note that using restarted GMRES may lead to superior performance but we do not explore that in this article. As a stopping tolerance for the iterative solvers, we use $10^{-5}$, following \cite{yurkin2007discrete}.

\begin{table*}[t!]
\centering
	\begin{tabular}{c c l   ||     l r     |      l r   ||      r       |       l r        |         l r}
	\multirow{2}{*}{$\mu$} &\multirow{2}{*}{$x$}& \#Total & \multicolumn{4}{c||}{Unpreconditioned} & \multicolumn{5}{c}{Preconditioned}  \\
	      & & voxels &  \multicolumn{2}{c | }{GMRES} & \multicolumn{2}{c||}{BiCG-Stab} &  & \multicolumn{2}{c | }{GMRES} & \multicolumn{2}{c}{BiCG-Stab} \\
	      & &            & Its. & Solve(s) & Its. & Solve(s) & Build(s)&  Its. & Solve(s) & Its. & Solve(s) \\
	      	      \hline\hline
	   1.2 & 10 & $2.7\times10^3$ & 8 & 0.38 & 8 & 0.10 & 0.29 & 6 & 0.39 & 7 & 0.08 \\
	   	 & 20 & $2.1\times10^4$ & 14 & 1.9 & 15 & 0.36 & 0.99 & 11 & 2.0 & 12 & 0.45 \\
		 & 30 & $6.9\times10^4$ & 28 & 7.3 & 33 & 1.9 & 2.4 & 27 & 8.0 & 34 & 3.1 \\
		 & 40 & $1.6\times10^5$ & 58 & 28 & 64 & 9.7 & 4.9 & 31 & 21 & 34 & 8.6 \\
		 & 60 & $5.5\times10^5$ & 189 & 72 & 247 & 120 & 17 & 32 & 29 & 36 & 31 \\ 
		 & 80 & $1.2\times10^6$ & 416 & 3,800 & 556 & 620 & 40 & 37 & 184 & 41 & 86 \\
		 &100 & $2.4\times10^6$ & 795 & 36,000 & 1062 & 2400 & 93 & 42 & 387 & 47 & 200 \\
		 \hline
	  1.4  & 10 & $3.5\times10^3$ & 10 & 0.45 & 11 & 0.071 & 0.63 & 7 & 0.47 & 8 & 0.072 \\
	  	& 20 & $3.5\times10^4$ & 41 & 4.6 & 47 & 1.4 & 1.6 & 27 & 4.2 & 39 & 1.9 \\
		& 30 & $1.1\times10^5$ & 158 & 61 & 233 & 20 & 3.3 & 35 & 15 & 43 & 6.1 \\
		& 40 & $2.5\times10^5 $ & 400 & 670 & 554 & 120 & 7.2 & 42 & 39 & 52 & 19 \\
		& 60 & $8.1\times10^5$ & 1533 & 27,000 & 2957 & 2100 & 24 & 64 & 180 & 98 & 120 \\
		& 80 & $1.9\times10^6$ & \tikzcircle{2pt} & \tikzcircle{2pt} & \tikzcircle{2pt} & \tikzcircle{2pt} & 81 & 107 & 750 & 234 & 750 \\
		&100 & $3.8\times10^6$ & \tikzcircle{2pt} & \tikzcircle{2pt} & \tikzcircle{2pt} & \tikzcircle{2pt} & 150 & 116 & 1800 & 217 & 1500 \\
		\hline
	1.6   & 10 & $6.7\times10^3$ & 16 & 0.75 & 17 & 0.13 & 0.43 & 11 & 0.76 & 12 & 0.15 \\
		& 20 & $4.5\times10^4$ & 85 & 11 & 111 & 4.1 & 1.6 & 29 & 5.6 & 33 & 2.2 \\
		& 30 & $1.6\times10^5$ & 573 & 850 & 1124 & 140 & 4.6 & 52 & 30 & 105 & 23 \\
		& 40 & $3.6\times10^5$ & 1339 & 9500 & 3460 & 1100 & 10 & 73 & 92 & 132 & 71 \\
		& 60 & $1.2\times10^6$ & \tikzcircle{2pt} & \tikzcircle{2pt} & \tikzcircle{2pt} & \tikzcircle{2pt} & 38 & 111 & 470 & 570 & 1100 \\
		& 80 & $2.9\times10^6$ & \tikzcircle{2pt} & \tikzcircle{2pt} & \tikzcircle{2pt} & \tikzcircle{2pt} & 110 & 195 & 2,700 & \tikzcircle{2pt} & \tikzcircle{2pt} \\
		& 100 & $5.9\times10^6$ & \tikzcircle{2pt} & \tikzcircle{2pt} & \tikzcircle{2pt} & \tikzcircle{2pt} & 290 & 247 & 12,000 & \tikzcircle{2pt} & \tikzcircle{2pt} \\
		\hline
	1.8   & 10 & $8.7\times10^3$    & 18 & 0.50 & 20 & 0.18 & 0.52 & 12 & 0.51 & 13 & 0.21 \\
		& 20 & $6.9\times10^4$ 	& 313 & 120 & 601 & 32 & 2.3  & 36 & 6.5 & 49 & 4.5 \\
		& 30 & $2.3\times10^5$ 	& 1,224 & 5,100 & 3,471 & 730 & 6.4 & 72 & 52 & 425 & 160 \\
		& 40 & $5.5\times10^5$ 	& \tikzcircle{2pt} & \tikzcircle{2pt} & \tikzcircle{2pt} & \tikzcircle{2pt} & 16 & 119 & 230 & \tikzcircle{2pt} & \tikzcircle{2pt}\\
		& 60 & $1.7\times10^6$ 	& \tikzcircle{2pt} & \tikzcircle{2pt} & \tikzcircle{2pt} & \tikzcircle{2pt} & 60 & 207 & 1,800 & \tikzcircle{2pt} & \tikzcircle{2pt} \\
		& 80 & $4.2\times10^6$ 	& \tikzcircle{2pt} & \tikzcircle{2pt} & \tikzcircle{2pt} & \tikzcircle{2pt} & 180 & 367 & 13,000 & \tikzcircle{2pt} & \tikzcircle{2pt} \\
		& 100 & $8.2\times10^6$ 	& \tikzcircle{2pt} & \tikzcircle{2pt} & \tikzcircle{2pt} & \tikzcircle{2pt} & 413 & \tikzcirclehol{2pt} & \tikzcirclehol{2pt} & \tikzcircle{2pt} & \tikzcircle{2pt} \\
		\hline
	2      & 10 & $1.1\times10^4$    & 20 & 0.56 & 24 & 0.24 & 0.61 & 13 & 0.57 & 13 & 0.25 \\
		& 20 & $8.5\times10^4$    & 502 & 350 & 1,313 & 91 & 2.7 & 46 & 11 & 139 & 19 \\
		& 30 & $3.2\times10^5$    & \tikzcircle{2pt} & \tikzcircle{2pt} & \tikzcircle{2pt} & \tikzcircle{2pt} & 8.5 & 114 & 120 & 706 & 320 \\
		& 40 & $7.3\times10^5$    & \tikzcircle{2pt} & \tikzcircle{2pt} & \tikzcircle{2pt} & \tikzcircle{2pt} & 38 & 191 & 630 & \tikzcircle{2pt} & \tikzcircle{2pt} \\
		& 60 & $2.4\times10^6$    & \tikzcircle{2pt} & \tikzcircle{2pt} & \tikzcircle{2pt} & \tikzcircle{2pt} & 97 & 407 & 7,500 & \tikzcircle{2pt} & \tikzcircle{2pt} \\
		& 80 & $5.9\times10^6$    & \tikzcircle{2pt} & \tikzcircle{2pt} & \tikzcircle{2pt} & \tikzcircle{2pt} & 280 & \tikzcirclehol{2pt}& \tikzcirclehol{2pt} & \tikzcircle{2pt} & \tikzcircle{2pt} \\
		&100 & $1.1\times10^7$    & \tikzcircle{2pt} & \tikzcircle{2pt} & \tikzcircle{2pt} & \tikzcircle{2pt} & 600 & \tikzcirclehol{2pt} & \tikzcirclehol{2pt} & \tikzcircle{2pt} & \tikzcircle{2pt} \\
	 \end{tabular}
	 \caption{Aspect ratio $L/a=0.1$. The symbol \tikzcircle{2pt} represents no convergence for GMRES within 2000 iterations and BiCG-Stab within 4000 iterations. The symbol \tikzcirclehol{2pt} signifies that the computer's memory limit was exceeded by GMRES's Krylov subspace.}
	 \label{tab:tenth}
\end{table*}
Table~\ref{tab:tenth} shows the iteration counts and timings for the hexagonal plate of aspect ratio $L/a = 0.1$. We are employing GMRES and BiCG-Stab as the iterative solvers and choose to cease the solves if convergence is not achieved within 2000 and 4000 iterations, respectively. One can observe that, with no preconditioning, the iteration count grows approximately as $\mathcal{O}(x^3)$ with both GMRES and BiCG-Stab, so that for large values of $\mu$, DDA simulations are infeasible, which motivates the use of a good preconditioner. We note further that BiCG-Stab is indeed faster than GMRES for many of the unpreconditioned simulations. 

\begin{figure}[t!]
	\centering
	\includegraphics[width=0.5\linewidth]{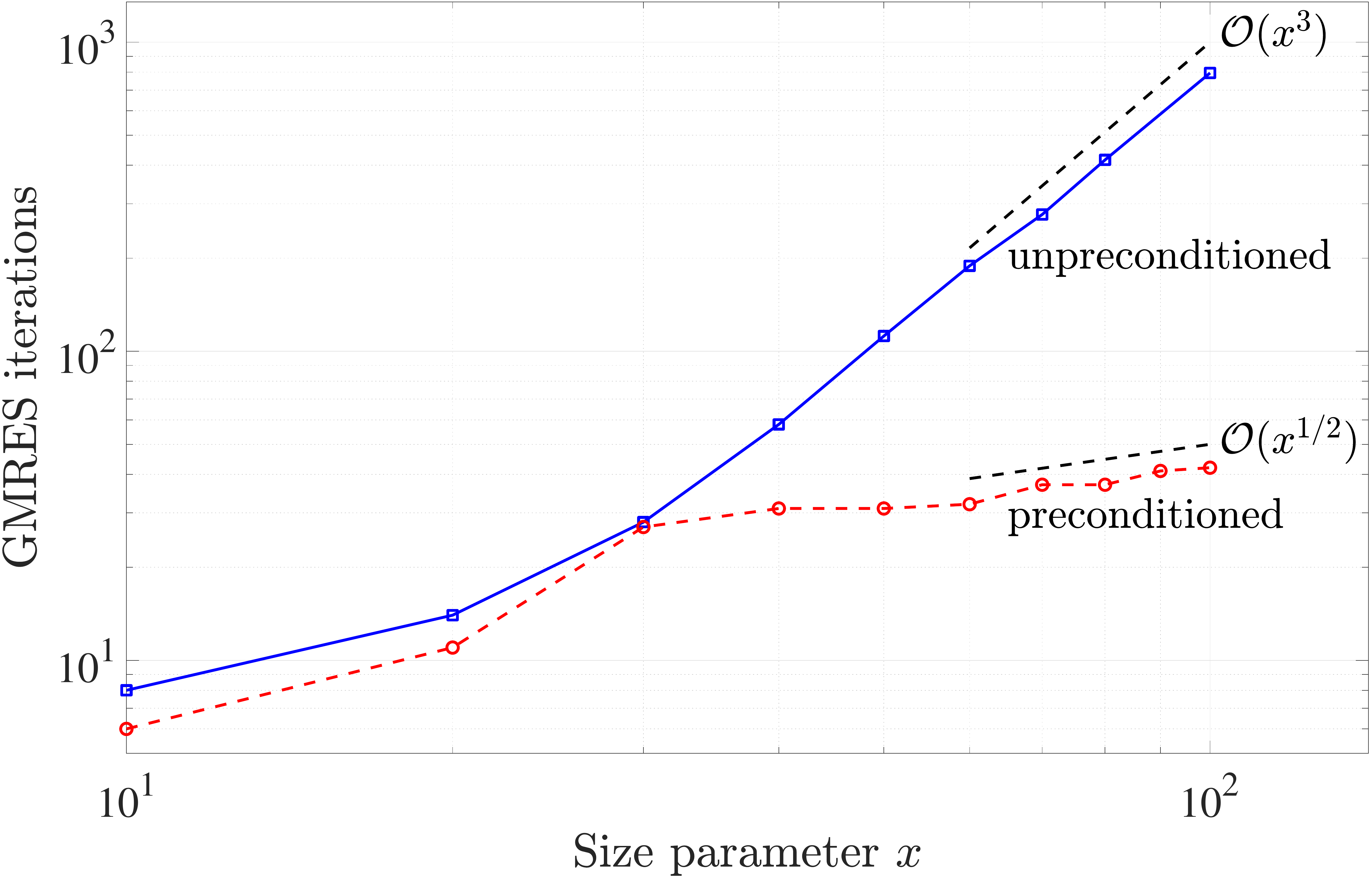}
	\caption{Iteration counts for GMRES ($\text{tol}=10^{-5}$) for the hexagonal plate with refractive index $\mu=1.2$, aspect ratio $L/a=0.1$, and size parameter ranging from 10 to 100. Without preconditioning, the iteration count grows slightly slower than $\mathcal{O}(x^3)$ whereas with the preconditioner, the growth is approximately $\mathcal{O}(x^{1/2})$.}
	\label{fig:GMRES_its}
\end{figure}
\begin{figure}[t!]
	\centering
	\includegraphics[width=0.5\linewidth]{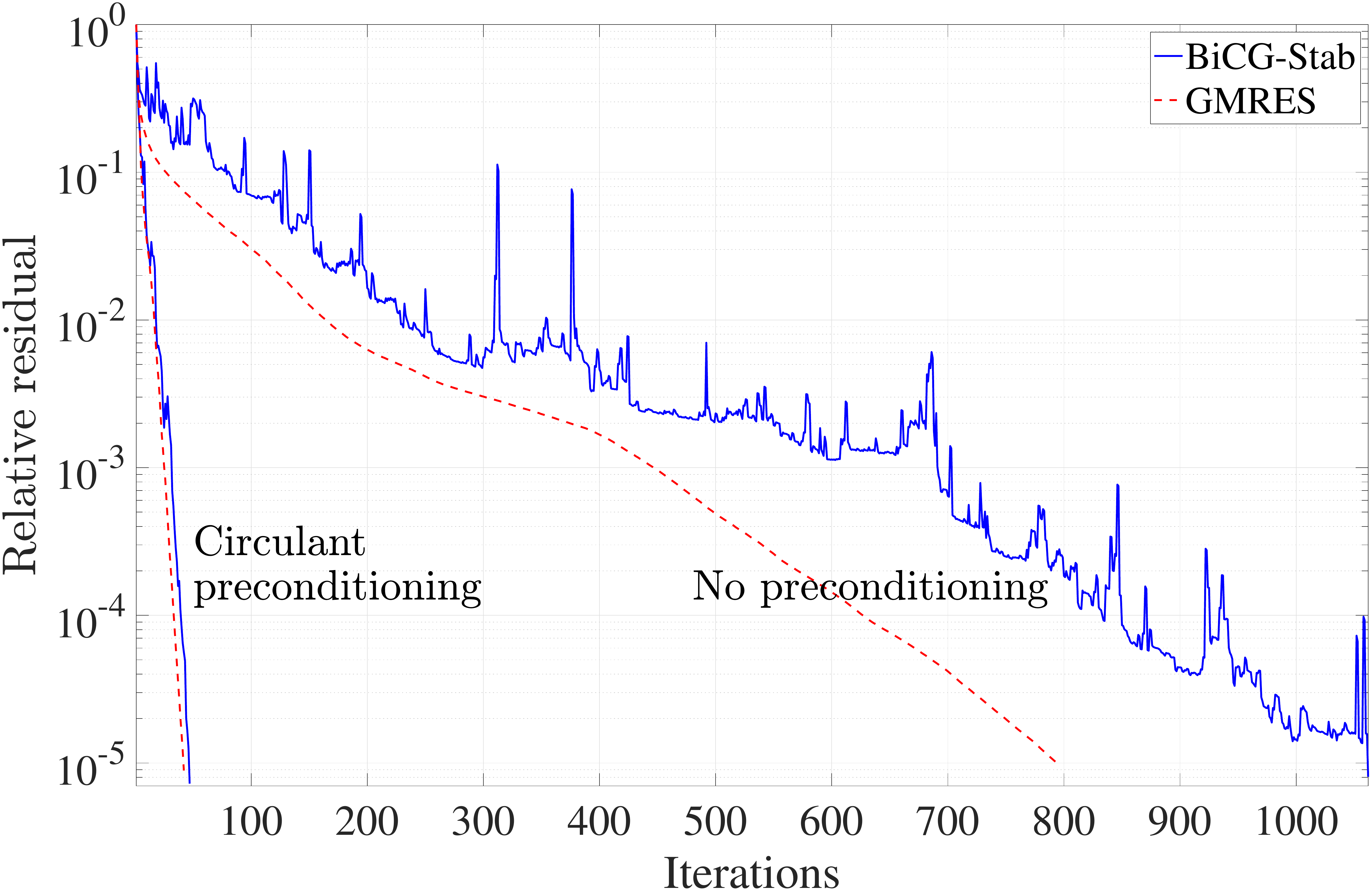}
	\caption{Convergence of BiCG-Stab and GMRES for a hexagonal plate of aspect ratio $L/a=0.1$, refractive index $\mu=1.2$, and size parameter $x=100$. For this example, there are $1.8\times10^6$ unknowns. The preconditioner took 92.6s to build and the solve times were the following: unpreconditioned BiCG-Stab - 1062s; preconditioned BiCG-Stab - 196s; unpreconditioned GMRES - 36,000s; preconditioned GMRES - 387s.}
	\label{fig:convergence}
\end{figure}
With the preconditioner, the iteration count grows much more slowly as the size parameter increases. For $\mu=1.2$, the iteration count growth is $\mathcal{O}(x^{1/2})$ (as shown in Figure~\ref{fig:GMRES_its}) whereas for larger $\mu$, the growth is closer to $\mathcal{O}(x)$, thereby permitting much larger size parameter simulations compared to without preconditioning. It is worthwhile to observe that BiCG-Stab yields faster preconditioned simulations than GMRES for $\mu=1.2, 1.4$ but for higher values of the refractive index $\mu$, GMRES proves more reliable. However, for $\mu=1.6, 1.8$ we see that the memory of the machine is exceeded by the Krylov subspace generated by GMRES at the largest size parameters. This suggests that exploring the use of GMRES with restarts would be worthwhile from a performance perspective. In any case, we observe that the preconditioner is providing an excellent improvement in the performance of iterative solvers. To illustrate this further, in Figure~\ref{fig:convergence} we present the convergence of the relative residual of GMRES and BiCG-Stab for $\mu=1.2, x=100, L/a=0.1$. 

\begin{figure}[t!]
\centering
\subfigure[Assembly and MVP timings]{
	\includegraphics[width=0.48\linewidth]{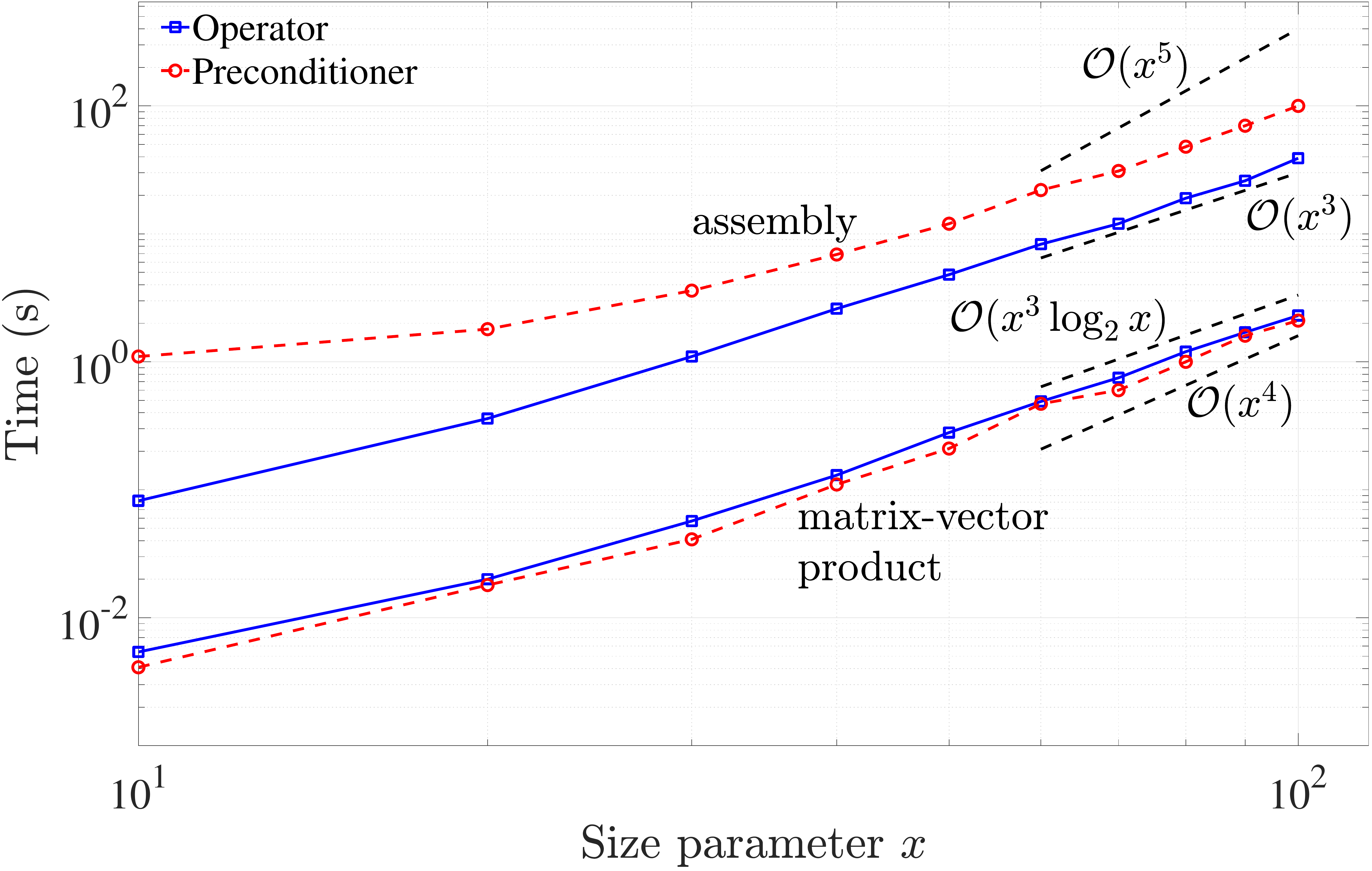}
	}
\subfigure[Memory consumption]{
	\includegraphics[width=0.48\linewidth]{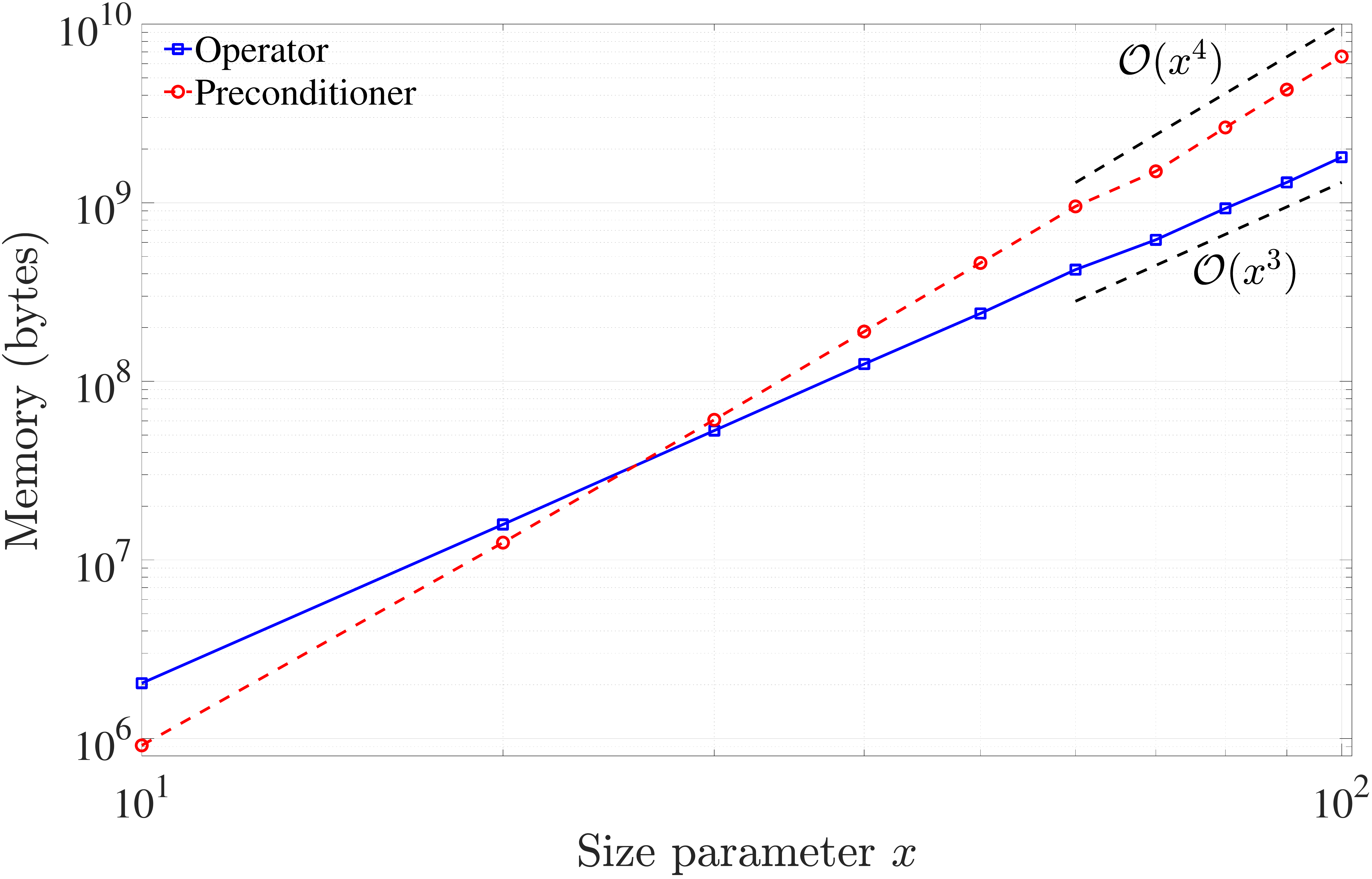}
	}
	\caption{Assembly times, matrix-vector product times, and memory consumption for the integral operator and preconditioner for the hexagonal plate of refractive index $\mu=1.2$ and aspect ratio $L/a=0.1$. Observe that the assembly of the preconditioner is much faster than our prediction of $\mathcal{O}(x^5)$.}
	\label{fig:MVP}
\end{figure}
In terms of timings, for the small size parameters, where little gain is achieved using the preconditioner, the simulation times are comparable between preconditioned and unpreconditioned solves, and the overhead of building the preconditioner is less than a second in all cases. For the larger size parameters, we observe the huge advantage gained by employing the preconditioner. For example, for $\mu=1.4, x=60$, the solve with BiCG-Stab without a preconditioner takes 35 minutes, whereas with the preconditioner the solve takes 2.5 minutes (including the 24s preconditioner build time) -- a factor of 14 speed up. So the small overhead time required to build the preconditioner is certainly worth it. 
\begin{table*}[ht!]
\centering
	\begin{tabular}{c c l   ||     l r     |      l r   ||      r       |       l r        |         l r}
	\multirow{2}{*}{$\mu$} &\multirow{2}{*}{$x$}& \#Total & \multicolumn{4}{c}{Unpreconditioned} & \multicolumn{5}{c}{Preconditioned}  \\
	      & & voxels &  \multicolumn{2}{c}{GMRES} & \multicolumn{2}{c}{BiCG-Stab} &  & \multicolumn{2}{c}{GMRES} & \multicolumn{2}{c}{BiCG-Stab} \\
	      & &            & Its. & Solve & Its. & Solve & Build &  Its. & Solve & Its. & Solve \\
	      	      \hline\hline
	   1.2 & 10 & $5.3\times10^3$ & 11 & 0.49 & 11 & 0.08 & 0.37 & 9 & 0.29 & 9 & 0.10 \\
		& 20 & $4.1\times10^4$ & 27 & 2.5 & 29 & 1.1 & 1.3 & 23 & 2.80 & 32 & 1.93 \\
		& 30 & $1.4\times10^5$ & 65 & 19 & 76 & 9.0 & 3.8 & 29 & 10.7 & 36 & 6.66 \\
		& 40 &  $3.1\times10^5$ & 120 & 100 & 149 & 37 & 8.4 & 35 & 30.3 & 38 & 16.8 \\
		& 60 & $1.1\times10^6$ & 308 & 1700 & 391 & 370 & 40 & 41 & 136 & 46 & 91.8 \\
		& 80 & $2.5\times10^6$ & 707 & 20,000 & 906 & 2,100 &130 & 55 & 610 & 72 & 630 \\
		& 100 & $4.8\times10^6$ & $\tikzcirclehol{2pt}$ & $\tikzcirclehol{2pt}$ & $\tikzcircle{2pt}$ & $\tikzcircle{2pt}$ & 290 & 61 & 2,500 & 79 & 2,800 \\
	\hline
	1.4 & 10 & $8.8\times10^3$ & 22 & 0.48 & 25 & 0.20 & 0.60 & 17 & 0.58 & 21 & 0.30 \\
	      & 20 & $6.3\times10^4$ & 116 & 20 & 156 & 7.6 & 1.9 & 31 & 5.1 & 46 & 3.4 \\
	      & 30 & $2.0\times10^5$ & 393 & 510 & 563 & 100 & 5.8 & 45 & 27 & 63 & 19 \\
	      & 40 & $5.0\times10^5$ & 852 & 5,400 & 1,692 & 760 & 17 & 60 & 89 & 96 & 82 \\
	      & 60 & $1.7\times 10^6$ & \tikzcircle{2pt} & \tikzcircle{2pt} & \tikzcircle{2pt} & \tikzcircle{2pt} & 74 & 95 & 680 & 159 & 670 \\
	      & 80 & $4.0\times10^6$ & \tikzcircle{2pt} & \tikzcircle{2pt}& \tikzcircle{2pt} & \tikzcircle{2pt} & 230 & 149 & 6,100 & 388 & 8,500 \\
	      & 100 &  $7.7\times10^6$ & \tikzcircle{2pt} & \tikzcircle{2pt} & \tikzcircle{2pt} & \tikzcircle{2pt} & 520 & \tikzcirclehol{2pt} & $\tikzcirclehol{2pt}$ & 956 & 19,000 \\
	 \hline
	 1.6 & 10 & $1.1\times10^4$ & 34 & 0.80 & 41 & 0.40 & 0.50 & 20 & 0.76 & 26 & 0.43 \\
	       & 20 & $9.0\times10^4$ & 384 & 230 & 641 & 44 & 2.5 & 42 & 9.7 & 58 & 6.8 \\
	       & 30 & $3.1\times10^5$ & 1,148 & 6,000 & 2,730 & 670 & 9.4 & 81 & 74 & 236 & 110 \\
	       & 40 & $7.2\times10^5$ & \tikzcircle{2pt} & \tikzcircle{2pt} & \tikzcircle{2pt} & \tikzcircle{2pt} & 26 & 114 & 300 & 330 & 411 \\
	       & 60 & $2.5\times10^6$ & \tikzcircle{2pt} & \tikzcircle{2pt} & \tikzcircle{2pt} & \tikzcircle{2pt} & 130 & 200 & 4,400 & 3,054 & 23,000 \\
	       & 80 & $5.9\times10^6$ & \tikzcircle{2pt} & \tikzcircle{2pt} & \tikzcircle{2pt} & \tikzcircle{2pt} & 362 & \tikzcirclehol{2pt} & \tikzcirclehol{2pt} & \tikzcircle{2pt} & \tikzcircle{2pt} \\
	 \hline
	 1.8 & 10 & $1.7\times10^4$ & 79 & 3.2 & 131 & 1.8 & 0.93 & 25 & 1.6 & 33 & 0.92 \\
	       & 20 & $1.4\times 10^5$ & 802 & 1,400 & 2,333 & 230 & 3.9 & 61 & 23 & 178 & 31 \\
	       & 30 & $4.4\times 10^5$ & \tikzcircle{2pt} & \tikzcircle{2pt} & \tikzcircle{2pt} & \tikzcircle{2pt} & 14 & 127 & 205 & \tikzcircle{2pt} & \tikzcircle{2pt} \\
	       & 40 & $1.1\times10^6$ & \tikzcircle{2pt} & \tikzcircle{2pt} & \tikzcircle{2pt} & \tikzcircle{2pt} & 41 & 222 & 1,300 & \tikzcircle{2pt} & \tikzcircle{2pt} \\
	       & 60 & $3.5\times10^6$ & \tikzcircle{2pt} & \tikzcircle{2pt} & \tikzcircle{2pt} & \tikzcircle{2pt} & 190 & 461 & 20,000 & \tikzcircle{2pt} & \tikzcircle{2pt} \\
	  \hline
	  2 & 10 & $2.1\times10^4$ & 133 & 9.1 & 256 & 4.2 & 0.78 & 30 & 1.8 & 42 & 1.2 \\
	     & 20 & $1.8\times10^5$ & 1,268 & 4,400 & \tikzcircle{2pt}  & \tikzcircle{2pt}  & 5.3 & 111 & 75 & \tikzcircle{2pt}  & \tikzcircle{2pt}  \\
	     & 30 & $6.0\times10^5$ & \tikzcircle{2pt} & \tikzcircle{2pt} & \tikzcircle{2pt} & \tikzcircle{2pt} & 23 & 216 & 620 &  \tikzcircle{2pt} &  \tikzcircle{2pt} \\
	     & 40 & $1.5\times10^6$ & \tikzcircle{2pt} & \tikzcircle{2pt} &  \tikzcircle{2pt} & \tikzcircle{2pt} & 65 & 443 & 5,740 &  \tikzcircle{2pt} &  \tikzcircle{2pt} \\
	     & 60 & $4.8\times10^6$ & \tikzcircle{2pt} & \tikzcircle{2pt} &  \tikzcircle{2pt} & \tikzcircle{2pt} & 280 & \tikzcirclehol{2pt} & \tikzcirclehol{2pt} &  \tikzcircle{2pt} &  \tikzcircle{2pt} \\
	      		 \end{tabular}
	 \caption{Aspect ratio $L/a=0.2$. The symbol \tikzcircle{2pt} represents no convergence for GMRES within 2000 iterations and BiCG-Stab within 4000 iterations. The symbol \tikzcirclehol{2pt} signifies that the computer's memory limit was exceeded by GMRES's Krylov subspace.}
	 \label{tab:fifth}
\end{table*}

In Figure~\ref{fig:MVP} we present more details of the overhead required to use the preconditioner. Figure~\ref{fig:MVP}(a) compares the assembly times of the preconditioner and the integral operator $\textbf{G}$ for growing size parameter $x$. In terms of the assembly, the time for $\textbf{G}$ grows as $\mathcal{O}(x^3)$ as can be seen from Algorithm~\ref{alg:G}. The cost of assembling the preconditioner is slightly higher and appears to increase as $\mathcal{O}(x^3)$ also, which is contrary to our prediction of $\mathcal{O}(x^5)$ in Section~\ref{subsec:costings}. This is likely due to the fact that the assembly is parallelized and Matlab's matrix inversion routines are extremely efficient and so it takes a very large $x$ before the asymptotic range of $\mathcal{O}(x^5)$ is reached. In this range of $x$, the preconditioner takes approximately 2.5 times as long to assemble as the operator $\textbf{G}$ but we note that the total simulation time is dominated by the iterative solve, so this increased setup time is worth it as long as the iteration count is reduced sufficiently. 

Also in Fig.~\ref{fig:MVP}(a) are shown the times required to perform matrix-vector products with the preconditioner as well as with the operator $\mathbf{G}$. We observe that they are comparable and agree well with the costings provided in Section~\ref{subsec:costings}. Since they are comparable, this suggests that the break even point in using the preconditioner is to reduce the iteration count to approximately half. That is, if the iteration count for the preconditioned solve is smaller than half that of the unpreconditioned solve, then employing the preconditioner is worthwhile. Indeed, we see in Table~\ref{tab:tenth} that this is indeed the case for the majority of the parameter combinations. 

Finally, in Fig.~\ref{fig:MVP}(b) we compare the memory required to store the preconditioner and integral operator. The memory required to store the preconditioner grows as $\mathcal{O}(x^4)$ compared to the $\mathcal{O}(x^3)$ required for the operator, with the preconditioner being more expensive for $x\geq 30$. This increased memory consumption is not problematic for the problems looked at here, however for higher resolution simulations and/or larger scatterers, this may prove a bottleneck. It is likely that some compression of the preconditioner is possible, as was seen in \cite{groth2019circulant} for the 1-level circulant preconditioner.

The final results presented are for the same scattering setup but now with an aspect ratio of $L/a = 0.2$, in Table~\ref{tab:fifth}. 
This scatterer is twice as large as the previous one so we would expect that the iteration counts for the unpreconditioned solves are even higher. Indeed we find this is the case - unpreconditioned solves require roughly twice the number of iterations compared to the $L/a=0.1$ example. The preconditioned solves also require more iterations but the increase is slightly less severe. The most noteworthy aspect is that now the timings for the preconditioned solves with GMRES are more comparable to those with BiCG-Stab, and certainly more reliable. For $\mu=1.8, 2$, BiCG-Stab struggles to converge except for at the lowest values of $x$. Again we see that for the most challenging problems, the memory consumption of GMRES without restarts is prohibitive, again motivating future numerical experimentation with restarted GMRES.

\section{Conclusions and future work}
\label{sec:conc}
In this paper, we have presented the first application of a multi-level circulant preconditioner to electromagnetic scattering simulations with the discrete dipole approximation. Indeed, we believe that this is the first presentation of a viable preconditioner of any kind for ice crystal simulations within the DDA literature. 

In particular, we applied the so-called ``optimal'' multi-level preconditioner of Chan and Olkin~\cite{chan1994circulant} in the simulation of scattering by homogeneous hexagonal plates of various size parameters and refractive indices. Via a consideration of the symmetrical block-Toeplitz Toeplitz-block structure of the voxel-discretized dyadic Green's function $\textbf{G}$, we provided costings for the assembly and application of the 2-level circulant preconditioner. These costings suggest that the number of flops required for the assembly of the preconditioner scales with the size parameter $x$ as $\mathcal{O}(x^5)$, compared to the standard DDA cost of $\mathcal{O}(x^3)$. However, it was seen in the numerical experiments that the cost scaling of the preconditioner is milder than this, close to $\mathcal{O}(x^3)$ for the size parameters considered. So that the preconditioner is approximately 2.5 times more expensive than the assembly of $\textbf{G}$, which is of the order of seconds or minutes. 

Further, we observed that the cost of applying the preconditioner is almost the same as the cost of an MVP with $\textbf{G}$, suggesting that a reduction in iteration count by a factor of two is the break-even point. For the vast majority of the parameter combinations considered, a reduction factor far greater than two was achieved. In fact, the iteration count appears to grow as $\mathcal{O}(x)$ (or even more mildy) compared to the $\mathcal{O}(x^3)$ growth seen for unpreconditioned solves. Hence, for larger size parameters, the reduction in solve time achieved by using the preconditioner is greatest. For some parameter combinations, the preconditioned solves were up to 15 times faster. More remarkably, for large size parameter and large refractive index scatterers, the preconditioner enables previously infeasible problems to become tractable, thereby enabling a wider applicability of DDA.

This work therefore has shown that circulant preconditioners for DDA simulations can be extremely effective. The scatterers considered here, although limited in their variety, are already of importance in atmospheric physics applications. For more general scattering setups, further experimentation and, potentially, development is required, however it seems probable that a circulant preconditioning strategy will prove helpful. We conclude by discussing some directions for future development of this work. 

Recall that in Section~\ref{ss:algorithms} a choice was made in the construction of the 2-level preconditioner. In particular, we created a constant diagonal matrix $\tilde{\alpha}^{-1}\textbf{I}$ to replace the matrix $\boldsymbol{\alpha}^{-1}$. This was done by simply using the value of $\alpha$ for ``ice'' for the air voxels also. Such an approximation step, however, may not be necessary. Instead, investigation of ``Circulant-plus-diagonal'' preconditioners, such as in ~\cite{ng2010approximate}, may prove fruitful. Or potentially a more sophisticated point-circulant approximation such as was considered in \cite{remis2012circulant} for one-dimensional problems.

Another choice was made in this article. Namely, we chose to perform the circulant approximation in the $x$- and $y$-directions since, for the hexagonal plates considered here, these are the largest dimensions. However, for hexagonal prisms of larger aspect ratio, superior results may be gained by choosing the longitudinal ($z$) axis as one of the circulant approximation directions. 

Finally, we remark upon the memory consumption of the preconditioner and the GMRES Krylov subspace. The memory consumption of the preconditioner scales as $\mathcal{O}(x^4)$ which was not problematic for the simulations performed here. However, for higher resolution and/or larger size parameter simulations, it may be desirable to compress the preconditioner in some way. This was seen to be achievable for the 1-level preconditioner in \cite{groth2019circulant} so it likely that some compression can also be obtained in the 2-level case, which may in turn lead to faster assembly times. 

The iterative solves with GMRES were seen to be more reliable than those with BiCG-Stab, however much more memory intensive due the storage of the entire Krylov subspace. The entire Krylov subspace was retained here since it guarantees convergence of solver, but it is not necessary. Experimentation with GMRES with (deflated) restarts (e.g., \cite{morgan2002gmres}) would be useful to determine a reliable, fast, and memory efficient strategy for performing the preconditioned solves with GMRES.

\section*{Funding}
This work was supported by a grant from Skoltech as part of the Skoltech- MIT Next Generation Program, and the Design for Manufacturability (DFM) Methods, PDK Extensions, and Tools for Photonic Systems project sponsored by AIM Photonics.

\bibliography{MIT_papers}
\bibliographystyle{ieeetr}

\end{document}